\documentclass[a4paper,11pt]{article}
\usepackage[dvipdfmx]{graphicx}
\usepackage{amsfonts}
\usepackage{physics,amsmath}
\usepackage{amssymb}
\usepackage{mathtools}
\usepackage{colortbl}
\usepackage{cite}
\usepackage{here}
\usepackage{hyperref}
\usepackage{mathrsfs}
\usepackage{algpseudocode}
\usepackage{xcolor}
\usepackage{ulem}
\usepackage{comment}
\numberwithin{equation}{section}
\usepackage[left=2cm,right=2cm,top=3.5cm,bottom=3.5cm]{geometry}

\allowdisplaybreaks[1]
\hypersetup{
	colorlinks=true,
	linkcolor=ceruleanblue,
	filecolor=ceruleanblue,      
	urlcolor=ceruleanblue,
	citecolor=ceruleanblue,
}
\definecolor{ceruleanblue}{rgb}{0.0, 0.2, 0.6}

\newcommand{\de}{{\rm d}}

\date{\today}

\begin{document}

\begin{flushright} {\footnotesize YITP-24-86, IPMU24-0030}  \end{flushright}

\begin{center}
\LARGE{\bf Bridging Dark Energy and Black Holes with EFT: Frame Transformation and Gravitational Wave Speed}
\\[1cm] 

\large{Shinji Mukohyama$^{\,\rm a, \rm b}$, Emeric Seraille$^{\,\rm a, \rm c}$, Kazufumi Takahashi$^{\,\rm a}$, \\ and Vicharit Yingcharoenrat$^{\,\rm b, \rm d}$}
\\[0.5cm]

\small{
\textit{$^{\rm a}$
Center for Gravitational Physics and Quantum Information, Yukawa Institute for Theoretical Physics, 
\\ Kyoto University, 606-8502, Kyoto, Japan}}
\vspace{.2cm}

\small{
\textit{$^{\rm b}$
Kavli Institute for the Physics and Mathematics of the Universe (WPI), The University of Tokyo Institutes for Advanced Study (UTIAS), The University of Tokyo, Kashiwa, Chiba 277-8583, Japan}}
\vspace{.2cm}

\small{
\textit{$^{\rm c}$
\'{E}cole Normale Sup\'{e}rieure, 45 Rue d’Ulm, F-75230 Paris, France}}
\vspace{.2cm}

\small{
\textit{$^{\rm d}$
High Energy Physics Research Unit, Department of Physics, Faculty of Science, Chulalongkorn University, Pathumwan, Bangkok 10330, Thailand}}
\vspace{.2cm}
\end{center}

\vspace{0.3cm} 

\begin{abstract}\normalsize
Typically, constraints on parameters of the effective field theory (EFT) of dark energy have been obtained in the Jordan frame, where matter fields are minimally coupled to gravity. To connect these constraints with those of the EFT of black hole perturbations with a timelike scalar profile, it is necessary to perform a frame transformation on the EFT in general. In this paper, we study the conformal/disformal transformation of EFT parameters on an arbitrary background. 
Furthermore, we explore the effect of an EFT operator $M_6(r) \bar{\sigma}^{\mu}_{\nu} \delta K^{\nu}_{\alpha} \delta K^{\alpha}_{\mu}$, which is elusive to the LIGO/Virgo bound on gravitational-wave speed, on the dynamics of odd-parity black hole perturbations. Intriguingly, a deviation from luminal propagation shows up only in the vicinity of the black hole, and the speeds of perturbations in the radial and angular directions are different in general due to the traceless part $\bar{\sigma}^\mu_\nu$ of the background extrinsic curvature. This study establishes an important link between cosmological constraints and those obtained in the black hole regime.
\end{abstract}

\vspace{0.3cm} 

\vspace{2cm}

\newpage
{
\hypersetup{linkcolor=black}
\tableofcontents
}

\flushbottom

\vspace{1cm}

%%%% -- Introduction -- %%%%%%%%%%%%%%%%%%%%%%%%%%%%%%%%%%%%%%%%%%%%%%%%%%%%%%%%

%%%%%%%%%%%%%%%%%%%%%%%%%%%%%%%%%%%%%%%%%%%%%%%%%%%%%%%%%%%%%%%%%%%%%%%%%%%%%
\section{Introduction}
Theories beyond general relativity (GR) have been extensively studied in the past decades~\cite{Koyama:2015vza,Ferreira:2019xrr,Arai:2022ilw}.
One typically introduces those modifications in order to address several mysteries in our Universe such as dark matter, dark energy (DE), and inflation, that GR could not fully explain.
Another motivation is that, on short scales, studying those modified gravity models may shed light on a consistent theory of quantum gravity.
Additionally, since the deviations from GR due to modified-gravity effects can in principle be constrained by observations/experiments, the study of modified gravity provides an efficient test of the validity of GR on scales relevant to the observations/experiments.

According to Lovelock's theorem~\cite{Lovelock:1971yv,Lovelock:1972vz}, in four space-time dimensions, the only possible covariant tensor~$E_{\mu\nu}$ that satisfies $\nabla^{\mu}E_{\mu\nu}=0$ as an identity and that is constructed only from a metric~$g_{\mu\nu}$ and its up-to-second-order derivatives is a linear combination of the Einstein tensor and the metric itself. Therefore, in order to modify GR, we need to relax at least one of the assumptions of the Lovelock's theorem. A simplest modification is to add a single scalar field, providing scalar-tensor theories. 
Traditional examples of such theories include the Brans-Dicke~\cite{Brans:1961sx} and the k-essence~\cite{Armendariz-Picon:2000nqq,Armendariz-Picon:2000ulo} theories.
In 1974, the most general class of scalar-tensor theories with covariant second-order Euler-Lagrange equations in a four-dimensional spacetime was formulated by Horndeski~\cite{Horndeski:1974wa} and was later independently rediscovered in the context of generalized Galileons~\cite{Deffayet:2011gz,Kobayashi:2011nu}.
Nevertheless, one can extend the Horndeski theories in a way that the Euler-Lagrange equations contain higher derivatives without the Ostrogradsky ghosts by imposing the so-called degeneracy conditions~\cite{Langlois:2015cwa,Motohashi:2016ftl}. Such scalar-tensor theories are called beyond Horndeski~\cite{Gleyzes:2014dya} and degenerate higher-order scalar-tensor (DHOST) theories~\cite{Langlois:2015cwa,Crisostomi:2016czh,Motohashi:2016ftl,BenAchour:2016fzp} (see \cite{Langlois:2018dxi,Kobayashi:2019hrl} for comprehensive reviews).
An extension of DHOST theories was achieved in the U-DHOST theories~\cite{DeFelice:2018ewo,DeFelice:2021hps,DeFelice:2022xvq}, which accommodate the so-called shadowy mode when the scalar field profile is timelike.
Another systematic way to extend the Horndeski theories is to perform an invertible (conformal or) disformal transformation~\cite{Bekenstein:1992pj,Bruneton:2007si} on the Horndeski action~\cite{Bettoni:2013diz,Zumalacarregui:2013pma}.\footnote{A non-invertible transformation can also generate a class of DHOST theories~\cite{Takahashi:2017pje,Langlois:2018jdg}, though it does not accommodate viable cosmology.}
Furthermore, applying a higher-derivative generalization of the disformal transformations~\cite{Takahashi:2021ttd,Takahashi:2023vva} to Horndeski theories and U-DHOST theories, one arrives at the generalized disformal Horndeski~\cite{Takahashi:2022mew} and generalized disformal unitary-degenerate theories~\cite{Takahashi:2023jro}, respectively.
(See also \cite{Takahashi:2022mew,Domenech:2023ryc,Tahara:2023pyg,Babichev:2024eoh} for yet further generalizations of invertible transformations.)
Although such generalized disformal transformations can spoil the consistency of matter coupling, there is a non-trivial subclass where this problem can be circumvented at least in the unitary gauge~\cite{Takahashi:2022mew,Naruko:2022vuh,Takahashi:2022ctx,Ikeda:2023ntu,Takahashi:2023vva}.

All such covariant scalar-tensor theories should in principle be described in a universal way by an effective field theory (EFT), given the symmetry breaking pattern and the background configuration. 
On Minkowski and de Sitter background, an EFT of scalar-tensor theories, known as ghost condensation, was first formulated in \cite{Arkani-Hamed:2003pdi,Arkani-Hamed:2003juy}. 
Such an EFT was constructed using the assumption that the time diffeomorphism invariance is spontaneously broken by the timelike gradient of a scalar field.
With this assumption, the EFT action constructed in the unitary gauge is invariant under the spatial diffeomorphism.
A similar approach was applied to cosmology where the background metric is assumed to be a Friedmann-Lema{\^i}tre-Robertson-Walker (FLRW) metric, which leads to the EFT of inflation/DE~\cite{Cheung:2007st,Gubitosi:2012hu}.
It is possible that an EFT describing the cosmological scale breaks down in the vicinity of a black hole (BH). 
Nevertheless, it remains possible that a single EFT describes both regimes, allowing in principle to obtain information about DE from astrophysical BHs.\footnote{For further discussions on potential methods to probe dark energy through black hole ringdown, see \cite{Noller:2019chl} and references therein.}
For this purpose, an EFT of perturbations on an arbitrary (including BH) background was systematically built in \cite{Mukohyama:2022enj}, respecting the timelike nature of the scalar field.\footnote{Complementarily to this work, an EFT of perturbations around a static and spherically symmetric BH background with a spacelike scalar profile was constructed in \cite{Franciolini:2018uyq}, and then extended to a slowly rotating BH in \cite{Hui:2021cpm}.}
(See also \cite{Khoury:2022zor} for the shift-symmetric EFT of BH perturbations with a timelike scalar profile.)
This EFT is particularly interesting as it would offer a possibility to make model-independent predictions of strong-gravity phenomena, which could be tested with gravitational wave (GW) observations by LIGO-Virgo-KAGRA collaborations~\cite{LIGOScientific:2016aoc,LIGOScientific:2018mvr,LIGOScientific:2019lzm,KAGRA:2023pio}.
As a phenomenological application of the EFT, the generalized Regge-Wheeler (RW) equation for odd-parity perturbations about a static and spherically symmetric BH background was derived in \cite{Mukohyama:2022skk}, based on which the quasinormal mode (QNM) spectrum~\cite{Mukohyama:2023xyf,Konoplya:2023ppx}, graybody factors~\cite{Konoplya:2023ppx,Oshita:2024fzf}, and tidal Love numbers~\cite{Barura:2024uog} were obtained.
It should be noted that the EFTs of cosmological/BH perturbations have recently been extended to include vector-tensor theories~\cite{Aoki:2021wew,Aoki:2023bmz}.

In the context of BH perturbations without matter fields, there is no notion of the Jordan frame, and hence one would choose a frame where the gravitational action has a simple form.
Indeed, the analyses in \cite{Mukohyama:2022skk,Mukohyama:2023xyf,Konoplya:2023ppx,Barura:2024uog} were carried out in a frame where the coefficient in front of the Einstein-Hilbert term is a constant, which we shall call an {\it almost Einstein frame}.
(We call an almost Einstein frame the {\it Einstein frame} when the scalar field does not affect the kinetic term of the metric.)
On the other hand, when matter fields are present, it is practically convenient to work in the Jordan frame where the matter fields are minimally coupled to gravity.
There are many situations where matter fields should be take into account.
One of the examples is an extreme mass ratio inspiral where a stellar-mass compact object (e.g., a neutron star) is orbiting around a supermassive BH, which is one of the promising targets for GW astronomy with future space-based observatories such as LISA.
Also, provided that many constraints on the parameters of the  EFT of DE (see, e.g., \cite{Piazza:2013pua,Traykova:2019oyx,Peirone:2019yjs}) were usually obtained in the Jordan frame, in order to connect those constraints to those of the EFT of BH perturbations in an almost Einstein frame, one has to perform a frame transformation, or more specifically, a conformal/disformal transformation on the EFT coefficients.
(See \cite{Creminelli:2017sry,Creminelli:2018xsv,Creminelli:2019nok,Creminelli:2019kjy} for further constraints on the parameters of the EFT of DE using the luminal propagation of GW and the absence of GW decay at LIGO/Virgo scales.)
Along this line of thought, in this paper, we study the conformal/disformal transformation of the EFT parameters on a generic background with a timelike scalar profile. 
These results will pave the way for studying the dynamics of perturbations in the Jordan frame as well as for connecting the bounds on cosmological scales to those applicable in the BH regime.

Moreover, we discuss the propagation speed of GWs in the EFT of BH perturbations, extending the result of \cite{Mukohyama:2022skk}.
In particular, we focus on odd-parity perturbations about a static and spherically symmetric BH and study the effect of an EFT operator~$M_6(r) \bar{\sigma}^\mu_\nu \delta K^\nu_\alpha \delta K^\alpha_\mu$ with $\bar{\sigma}^\mu_\nu$ being the traceless part of the background extrinsic curvature, which is specific to an inhomogeneous background and hence is elusive to the LIGO/Virgo bound on GW speed.
We show that a deviation from luminal propagation shows up only in the vicinity of the BH, and the speeds of perturbations in the radial and angular directions are different in general due to the traceless nature of $\bar{\sigma}^\mu_\nu$.

The rest of this paper is organized as follows. In Section~\ref{EFT_black_hole_perturbation}, we briefly review the construction of the EFT on an arbitrary background spacetime with a timelike scalar profile. In Section~\ref{Section_EFT_catalog}, we provide the conformal/disformal transformation of the EFT coefficients.
In Section~\ref{Section_constrain_c_gw}, we analyze the dynamics of odd-parity perturbations on a static and spherically symmetric background, taking into account an EFT operator~$M_6(r) \bar{\sigma}^\mu_\nu \delta K^\nu_\alpha \delta K^\alpha_\mu$.
Finally, we summarize the paper and discuss future directions in Section~\ref{sec:conclusions}.

%%%%%%%%%%%%%%%%%%%%%%%%%%%%%%%%%%%%%%%%%%%%%%%%%%%%%%%%%%%%%%%%%%%%%%%%%%%%%%%%%%%%%%%%%%%%%%%%%%%%%%%%%%
\section{Overview of the EFT}
\label{EFT_black_hole_perturbation}
Here, we briefly discuss the formulation of the EFT of perturbations on an arbitrary background with a timelike scalar profile. (See \cite{Mukohyama:2022enj} for details and \cite{Khoury:2022zor} for a similar construction of a shift-symmetric EFT of scalar-tensor theories.)
The setup of this EFT is very similar to that of ghost condensation~\cite{Arkani-Hamed:2003pdi} and the so-called EFT of inflation/DE~\cite{Cheung:2007st,Gubitosi:2012hu}. 
The starting point is the assumption that the background of a scalar field~$\Phi$ is chosen to be timelike everywhere in a spacetime region of interest, which then spontaneously breaks the time diffeomorphism invariance. 
We denote the background of $\Phi$ as $\bar{\Phi}(\tau)$, with $\tau$ being the time coordinate in terms of which the scalar field is spatially uniform.
As a result, the EFT action is invariant under the 3d diffeomorphism, which is the residual symmetry realized on a constant-$\Phi$ hypersurface. 
In fact, the scalar field defines a preferred (constant-$\Phi$) slicing, whose unit timelike normal vector (i.e., such that $n_{\mu}n^{\mu}=-1$) is defined by
\begin{equation}\label{eq:normal_vector}
n_{\mu} \equiv -\frac{\nabla_{\mu}{\Phi}}{\sqrt{-X}} \rightarrow -\frac{\delta_{\mu}^{\tau}}{\sqrt{-g^{\tau \tau}}} \;.
\end{equation}
Here, $X$ is the kinetic term of the scalar field defined by $X \equiv \nabla^{\mu}\Phi\nabla_{\mu}\Phi$, which is assumed to be negative.\footnote{In order for our EFT to connect BH physics with that of dark energy, the background of $X$ must remain similar in both regimes.}
The arrow in Eq.~(\ref{eq:normal_vector}) represents taking the unitary gauge where $\delta\Phi \equiv \Phi - \bar{\Phi} = 0$.
Below, we will explain the formulation of the EFT of perturbations in the unitary gauge.

Let us introduce the Arnowitt-Deser-Misner (ADM) decomposition of the metric:
\begin{equation}
    {{\rm d}s}^{2}=-N^{2} \de \tau^2 +h_{ij}\bigl(\de x^{i} +N^{i}\de \tau\bigr)\bigl(\de x^{j} + N^{j}\de \tau\bigr)\;,
\end{equation}
where Latin indices refer to the spatial coordinates, $N=1/\sqrt{-g^{\tau \tau}}$ is the lapse function, $N^{i}=g^{i \tau} N^2$ is the shift vector, and $h_{ij}=g_{ij}$ is the spatial metric induced on a constant-$\Phi$ hypersurface. 
The extrinsic curvature can be computed using the projection tensor~$h_{\mu\nu} \equiv g_{\mu \nu} + n_{\mu} n_{\nu}$:
\begin{equation}
K_{\mu \nu}=h_{\mu}^{\rho}\nabla_{\rho}n_{\nu}\;.
\end{equation}
Alternatively, in terms of the ADM variables, we have
\begin{equation}
K_{ij}=\frac{1}{2N}\bigl(\dot{h}_{i j}-{\rm D}_{i}N_{j}-{\rm D}_{j}N_{i}\bigr) \;, \qquad K = h^{ij}K_{ij} \;,
\end{equation}
with a dot denoting the derivative with respect to $\tau$, $K$ the trace of the extrinsic curvature, and ${\rm D}_i$ the spatial covariant derivative associated with the induced metric~$h_{ij}$. 
Furthermore, one can straightforwardly calculate other geometrical quantities lying on the constant-$\Phi$ hypersurface, for instance, the 3d Ricci tensor~${}^{(3)}\!R_{ij}$. 
It is useful to note that the 4d Riemann tensor and those 3d quantities such as ${}^{(3)}\!R_{ij}$ and $K_{ij}$ are related by the Gauss-Codazzi relations.

In what follows, we express the most general EFT action in the unitary gauge, respecting the 3d diffeomorphism invariance.
First, we note that all 4d diffeomorphism invariant quantities can be included, for example, the 4d Ricci scalar or a scalar function made out of the 4d Riemann tensor and its derivatives. 
In addition, the EFT is allowed to contain scalar functions built out of 3d quantities such as the lapse function~$N$ (or equivalently $g^{\tau \tau}$), the extrinsic curvature, and the spatial Ricci tensor.
On top of these, the action can contain an explicit $\tau$-dependence since the time diffeomorphism invariance has been broken.
Notice that one does not need to consider the 4d Riemann tensor, as it can be written in terms of the spatial Ricci tensor and the extrinsic curvature, thanks to the Gauss relation.
Note also that the 3d Riemann tensor can also be omitted since it can be expressed by the Ricci tensor in 3d.
Therefore, the most general EFT action in the unitary gauge is given by
\begin{equation}
\label{General_Action}
    S=\int{{\rm d}^{4}x\sqrt{-g}\,L\bigl(g^{\tau \tau},K^\mu_{\nu},{}^{(3)}\!R^\mu_{\nu},\nabla_{\mu},\tau\bigr)}\;,
\end{equation}
where $L$ is an arbitrary scalar function. 
Having said that, the EFT action above is not useful for describing dynamics of perturbations.
In practice, we Taylor-expand such an action order by order in perturbations around a non-trivial background geometry.
For instance, the perturbations of $g^{\tau\tau}$, $K^\mu_\nu$, and ${}^{(3)}\!R^\mu_{\nu}$ are defined as
\begin{equation}\label{eq:pert_def}
    \delta g^{\tau \tau} \equiv g^{\tau \tau}-\bar{g}^{\tau \tau}(\tau, \vec{x}) \;, \qquad \delta K^\mu_\nu \equiv K^\mu_\nu - \bar{K}^\mu_\nu(\tau, \vec{x}) \;, \qquad \delta {}^{(3)}\!R^\mu_\nu \equiv {}^{(3)}\!R^\mu_\nu -{}^{(3)}\!\bar{R}^\mu_\nu(\tau, \vec{x}) \;,
\end{equation}
where a bar refers to the background value. Notice that the background values can depend on both spatial and temporal coordinates due to the inhomogeneities of the background metric. 
Then, expanding the action~(\ref{General_Action}) in terms of the perturbations leads to the EFT of perturbations. 
Since the construction of the EFT was already presented in detail in \cite{Mukohyama:2022enj,Mukohyama:2022skk}, here we only discuss terms that were omitted in the previous work.

From the action~(\ref{General_Action}), the Taylor expansion gives
\begin{align}\label{eq:EFT_extra}
    S=\int \de^{4}x\sqrt{-g}\bigg[\bar{L}+\bar{L}_{g^{\tau \tau}}\delta g^{\tau \tau}+\bar{L}_{K}\delta K+\bar{L}_{\sigma^{\mu}_{\nu}}\delta \sigma^{\mu}_{\nu}+\bar{L}_{{}^{(3)}\!R}\delta {}^{(3)}\!R+\bar{L}_{r^{\mu}_{\nu}}\delta r^{\mu}_{\nu} + \frac{1}{2}\bar{L}_{\sigma^{\mu}_{\alpha} \sigma^{\alpha}_{\nu}} \delta \sigma^\mu_\rho \delta \sigma^\rho_\nu +\cdots \bigg]\;,
\end{align}
where the ellipsis denotes terms that were already taken into account in the previous work as well as those of higher order in perturbations and/or derivatives.
Here, we have used the notation~$\bar{L}_Q \equiv (\partial L/\partial Q)_{\rm bkg}$, and the terms linear in perturbations are responsible for the background dynamics.
In the above action, we have defined the traceless part of $K^\mu_\nu$ and ${}^{(3)}\!R^\mu_\nu$ respectively as $\sigma^\mu_\nu\equiv K^\mu_\nu-K h^\mu_\nu/3$ and $r^\mu_\nu \equiv {}^{(3)}\!R^\mu_\nu-{}^{(3)}\!R h^\mu_\nu/3$.
Note that, on a flat FLRW background, both $\bar{\sigma}^\mu_\nu$ and $\bar{r}^\mu_\nu$ are vanishing. However, they are generally present on a BH background, so that one expects a significant change in the observables due to operators containing the background of $\sigma^\mu_\nu$ and $r^\mu_\nu$.
In \cite{Mukohyama:2022skk,Mukohyama:2023xyf}, the coefficient of $\delta \sigma^\mu_\rho \delta \sigma^\rho_\nu$ in Eq.~\eqref{eq:EFT_extra} was assumed to be $\bar{L}_{\sigma^{\mu}_{\alpha} \sigma^{\alpha}_{\nu}}\propto \delta^\nu_\mu$ as the traceless part would be suppressed by derivatives in general.
However, in this paper, we show that the traceless part of $\bar{L}_{\sigma^{\mu}_{\alpha} \sigma^{\alpha}_{\nu}}$ leads to an interesting phenomenology of odd-parity BH perturbations.
In particular, in the subsequent Sections, we discuss the case of $\bar{L}_{\sigma^{\mu}_{\alpha} \sigma^{\alpha}_{\nu}}\supset \bar{\sigma}^\nu_\mu$.
Note that there is no non-trivial traceless tensor on a homogeneous and isotropic background, and therefore it potentially distinguishes between observables in cosmological regimes and those in the vicinity of BH.

Before going further, let us comment on the consistency relations that the EFT should satisfy.
Such relations are imposed on the EFT coefficients or the Taylor-expansion parameters in Eq.~(\ref{eq:EFT_extra}), in order for the EFT to be invariant under the spatial diffeomorphism.
This is because each term of the Taylor expansion of the EFT~(\ref{General_Action}) breaks the 3d diffeomorphism invariance, as the background values of the EFT building blocks do not transform covariantly under the 3d diffeomorphism.
With this reasoning, in order to ensure that the 3d diffeomorphism invariance of the EFT is preserved, one necessarily imposes a set of consistency relations.
As explained in \cite{Mukohyama:2022enj,Mukohyama:2022skk}, a set of infinitely many consistency relations among the EFT coefficients can be obtained by applying the chain rule associated with the spatial derivatives to each term of the Taylor expansion.\footnote{See \cite{Aoki:2021wew,Aoki:2023bmz} for a similar procedure for constructing the EFT of vector theories.}
Note that applying the same process to $\tau$-derivatives does not provide non-trivial relations among the EFT coefficients, as they are automatically satisfied by choosing an appropriate $\tau$-dependence of the EFT.
Moreover, it was verified in \cite{Mukohyama:2022enj,Mukohyama:2022skk} that these consistency relations trivially hold in the covariant theories, for instance, Horndeski and (U-)DHOST theories, as they should. 

Let us now write down the EFT action up to the second order in perturbations as
\begin{align}
\label{EFT_action}
    S= & \int \de^{4}x\sqrt{-g}\bigg[\frac{M_{\star}^{2}}{2}f(x) R - \Lambda(x) -  c(x) g^{\tau \tau} -\beta(x) K - \alpha(x)^{\mu}_{\nu} \sigma^{\nu}_{\mu}- \gamma(x)^{\mu}_{\nu} r^{\nu}_{\mu} -\zeta(x) \bar{n}^{\mu}\partial_{\mu}g^{\tau \tau} \nonumber \\
   &+ \frac{1}{2}m_{2}^{4}(x)(\delta g^{\tau \tau})^{2} + \frac{1}{2} M_{1}^{3}(x)\delta g^{\tau \tau}\delta K+\frac{1}{2} M_{2}^{2}(x)\delta K^{2} 
   +\frac{1}{2} M_3^2(x) \delta K^\mu_\nu \delta K^\nu_\mu
   + \frac{1}{2} M_{4}(x) \delta K \delta {}^{(3)}\!R \nonumber \\
   &+\frac{1}{2} M_{5}(x)\delta K^{\mu}_{\nu} \delta {}^{(3)}\!R^{\nu}_{\mu} +\frac{1}{2} \tilde{M}_6^2(x)^\mu_\nu \delta K^\nu_\alpha \delta K^\alpha_\mu
   +\frac{1}{2} \mu_{1}^{2}(x)\delta g^{\tau \tau}\delta {}^{(3)}\!R +\frac{1}{2} \mu_{2}(x) \delta {}^{(3)}\!R^{2} \nonumber \\
  &+\frac{1}{2} \mu_{3}(x)\delta {}^{(3)}\!R^{\mu}_{\nu} \delta {}^{(3)}\!R^{\nu}_{\mu}+\frac{1}{2}{\lambda_{1}(x)}^{\nu}_{\mu}\delta g^{\tau \tau} \delta K^{\mu}_{\nu}+\frac{1}{2}{\lambda_{2}(x)}^{\nu}_{\mu}\delta g^{\tau \tau} \delta {}^{(3)}\!R^{\mu}_{\nu}+\frac{1}{2}{\lambda_{3}(x)}^{\nu}_{\mu}\delta K \delta K^{\mu}_{\nu} \nonumber \\
  &+\frac{1}{2}{\lambda_{4}(x)}^{\nu}_{\mu}\delta K \delta {}^{(3)}\!R^{\mu}_{\nu}+\frac{1}{2}{\lambda_{5}(x)}^{\nu}_{\mu}\delta {}^{(3)}\!R \delta K^{\mu}_{\nu}+\frac{1}{2}{\lambda_{6}(x)}^{\nu}_{\mu}\delta {}^{(3)}\!R \delta {}^{(3)}\!R^{\mu}_{\nu} + \frac{1}{2}\mathcal{M}_{1}^{2}(x) (\bar{n}^{\mu}\partial_{\mu}\delta g^{\tau \tau})^{2} \nonumber \\ 
  &+ \frac{1}{2}\mathcal{M}_{2}^{2}(x) \delta K \bar{n}^{\mu}\partial_{\mu}\delta g^{\tau \tau} +\frac{1}{2}\mathcal{M}_{3}^{2}(x) \bar{h}^{\mu \nu}\partial_{\mu}\delta g^{\tau \tau}\partial_{\nu}\delta g^{\tau \tau}
  + \cdots \bigg] \;,
\end{align}
where the ellipsis refers to higher-order operators. Note that $R$ denotes the 4d Ricci scalar without the total derivative term:
\begin{equation}\label{eq:Ricci_bdy}
    R \equiv {}^{(3)}\!R + K^\mu_\nu K^\nu_\mu - K^2 = \tilde{R} - 2 \nabla_{\mu}\bigl(K n^{\mu} - n^{\nu}\nabla_{\nu}n^{\mu} \bigr)\;,
\end{equation}
where $\tilde{R}$ is the standard 4d Ricci scalar.
As usual, the terms in the first line of Eq.~(\ref{EFT_action}) describe the background dynamics.
It should be noted that we have included a new operator with the coefficient~$\tilde{M}_6^2(x)^\mu_\nu$, which is assumed to be traceless,\footnote{In other words, the trace part has been absorbed into $M_3^2$. The new operator should also be taken into account when one derives the consistency relations.} coming from the traceless part of the Taylor coefficient~$\bar{L}_{\sigma^\mu_\alpha\sigma^\alpha_\nu}$ in Eq.~\eqref{eq:EFT_extra}.
Also, in the action~(\ref{EFT_action}), we did not include terms that contain a generic rank-4 tensor coefficient [e.g., $\lambda_7(x)^{\nu\beta}{}_{\mu\alpha}\delta K^\mu_\nu\delta K^\alpha_\beta$] for simplicity.

Let us comment on the non-minimal coupling function~$f(x)$ in front of the Ricci scalar.
In general, $f(x)$ is a function of background quantities such as $\tau$, $\bar{g}^{\tau\tau}$, $\bar{K}$, and ${}^{(3)}\!\bar{R}$.
In the present paper, however, by invoking the derivative expansion and restricting our considerations to lower-order terms, we assume $f(x)$ is a function only of $\tau$ and $\bar{g}^{\tau\tau}$.
Equivalently, in the 4d covariant language, we assume that the non-minimal coupling function is a function only of $\Phi$ and $X$, and ignore higher-derivative interactions with the kinetic term of the metric.
Indeed, we expect that those higher-derivative interactions would be suppressed by a very large energy scale, e.g., the Planck scale.\footnote{On the other hand, we keep other higher-derivative interactions in Eq.~\eqref{EFT_action}, which we expect to be suppressed by much lower energy scale associated with the physics of DE.}
We will shortly see that this assumption guarantees the existence of an almost Einstein frame, i.e., a frame where $f=1$.

%%%%%%%%%%%%%%%%%%%%%%%%%%%%%%%%%%%%%%%%%%%%%%%%%%%%%%%%%%%%%%%%%%%%%%%%
\section{Conformal/disformal transformation of the EFT}
\label{Section_EFT_catalog}

In this Section, we study the transformation of the EFT parameters, introduced in Eq.~(\ref{EFT_action}), under a conformal/disformal transformation of $g_{\mu\nu}$: 
\begin{equation}
\label{disf}
    \hat{g}_{\mu \nu}= f_{0}\bigl(\Phi,X\bigr) g_{\mu \nu} + f_{1}\bigl(\Phi,X\bigr) \partial_{\mu}\Phi\partial_{\nu}\Phi\;,
\end{equation}
where $f_0$ and $f_1$ are functions of $\Phi$ and $X$, with the scalar field~$\Phi$ unchanged.
We assume
    \begin{align}
    f_0>0\;, \qquad
    Y>0\;, \qquad
    \bigg(\frac{X}{Y}\bigg)_X\ne 0\;,
    \end{align}
with $Y\equiv f_0+Xf_1$ and a subscript~$X$ denoting the derivative with respect to $X$, so that the transformation is invertible (and hence it does not change the number of physical degrees of freedom~\cite{Domenech:2015tca,Takahashi:2017zgr}) and the metric signature is preserved~\cite{Bekenstein:1992pj,Bruneton:2007si,Bettoni:2013diz,Zumalacarregui:2013pma}.
It is useful to note that the class of Horndeski theories is closed under disformal transformations with $f_0=f_0(\Phi)$ and $f_1=f_1(\Phi)$~\cite{Bettoni:2013diz} (see also Appendix~\ref{HorndeskiConformal}), while the quadratic/cubic DHOST class is closed under the general disformal transformation~\eqref{disf}.

Note in passing that a conformal transformation, i.e., Eq.~\eqref{disf} with $f_1=0$, can be used to remove the non-minimal coupling~$f(x)$ in front of the Ricci scalar [see Eq.~\eqref{EFT_action}].
In other words, one can move to an almost Einstein frame by performing an appropriate conformal transformation.
Indeed, as mentioned at the end of the previous section, we are now assuming that $f(x)$ is a function only of $\Phi$ and $X$ in the 4d covariant language, and hence a conformal transformation with $f_0=f(x)$ removes the non-minimal coupling.
Note also that a conformal transformation does not change the speed of fluctuations whose effective metric coincides with $g_{\mu\nu}$, while this is not the case with a disformal transformation where $f_1\ne 0$.

Before proceeding further, let us discuss a general structure of the EFT Lagrangian before and after applying the transformation~(\ref{disf}).
We note that, under a general disformal transformation~(\ref{disf}), the set of EFT parameters is not closed in general.
To see this, let us consider a hatted-frame covariant Lagrangian of the form
\begin{align}
    \hat{\cal L}=\hat{\cal L}(\hat{g}_{\mu\nu},\hat{R}_{\mu\nu\lambda\sigma},\Phi,\partial_\mu\Phi,\hat{\nabla}_\mu\hat{\nabla}_\nu\Phi)\;,
    \label{L_seed}
\end{align}
which contains at most second derivatives acting on $g_{\mu\nu}$ and $\Phi$. 
Schematically, Eq.~(\ref{L_seed}) can be written as
    \begin{align}
    \hat{\cal L}=\hat{\cal L}(\hat{g},\partial \hat{g},\partial^2 \hat{g},\Phi,\partial\Phi,\partial^2\Phi)\;.
    \label{L_seed_schematic}
    \end{align}
We now perform the disformal transformation~\eqref{disf} on this Lagrangian.
Note that the disformal metric~$\hat{g}$ is a function only of $g$, $\Phi$, and $\partial\Phi$, and it does not depend on derivatives of the metric, which means that
    \begin{align}
    \begin{split}
    \partial \hat{g}&=\partial \hat{g}\bigl(g,\partial g,\Phi,\partial\Phi,\partial^2\Phi\bigr)\;, \\
    \partial^2 \hat{g}&=\partial^2 \hat{g}\bigl(g,\partial g,\partial^2 g,\Phi,\partial\Phi,\partial^2\Phi,\partial^3\Phi\bigr)\;.
    \end{split}\label{dhatg_schematic}
    \end{align}
Hence, the Lagrangian in the unhatted frame has the following schematic form:
    \begin{align}
    {\cal L}={\cal L}(g,\partial g,\partial^2 g,\Phi,\partial\Phi,\partial^2\Phi,\partial^3\Phi)\;.
    \label{L_disf}
    \end{align}
The fact that there is an additional term $\partial^3\Phi$ in the Lagrangian after the disformal transformation implies that one needs new EFT building blocks, that can accommodate the term $\partial^3\Phi$.\footnote{In the case of quadratic/cubic DHOST theories, the second derivative acting on metric variables [symbolically denoted by $\partial^2 g$ in Eq.~\eqref{L_seed_schematic}] can be removed via integration by parts. Therefore, in this case, the Lagrangian is stable under the general disformal transformation~\eqref{disf} and one does not need to introduce new EFT building blocks.}
Nevertheless, even if we perform a second disformal transformation on the Lagrangian~(\ref{L_disf}), we do not get any higher derivatives like $\partial^3g$ or $\partial^4\Phi$.
This is simply because the transformation of $\partial^2 g$ contains at most the term~$\partial^2 g$ and $\partial^3 \Phi$, as shown in Eq.~\eqref{dhatg_schematic}.
Likewise, the schematic form of the Lagrangian remains the same even if we perform more than two disformal transformations.\footnote{In a sense, this is trivial as a functional composition of two (or more) conformal/disformal transformations of the form~\eqref{disf} is again a conformal/disformal transformation of the same form~\cite{Takahashi:2021ttd}.}

For demonstrative purposes, in what follows, we focus on the conformal or disformal transformation, assuming that the transformation functions~$f_0$ and $f_1$ depend only on $\Phi$. This can be straightforwardly extended to the cases where the conformal/disformal factors are $X$-dependent. In Appendix~\ref{Section_general_disformal}, we provide the basic ingredients to compute the transformation of the EFT parameters when $f_0$ and $f_1$ are functions of $X$ as well as $\Phi$.

\subsection{\texorpdfstring{$X$-independent conformal transformation}{X-independent conformal transformation}}
\label{subsectiontonlyconf}
In this Subsection, we consider a pure conformal transformation with $f_0=f_0(\Phi)$ of the EFT coefficients introduced in Eq.~(\ref{EFT_action}), i.e.,
\begin{equation}
    \hat{g}_{\mu \nu}= f_{0}(\Phi) g_{\mu \nu}\;.
    \label{tonlyconf}
\end{equation}
Note that, in the unitary gauge, $f_0$ is a function only of $\tau$.
For demonstration purposes, we omit the terms with $\zeta$, ${\cal M}_1$, ${\cal M}_2$, ${\cal M}_3$ in what follows.
Indeed, as we will see shortly, these terms are not necessary for the EFT to be closed under the pure conformal transformation~\eqref{tonlyconf}.\footnote{When the conformal factor has an $X$-dependence, one has to keep those terms to make the EFT closed under the transformation. (See also Appendix~\ref{Section_general_disformal}.)}
Also, regarding the newly-introduced term with the coefficient~$\tilde{M}_6^2(x)^\mu_\nu$, we assume $\tilde{M}_6^2(x)^\mu_\nu=M_6(x)\bar{\sigma}^\mu_\nu$ for concreteness.

For convenience, we give the transformations of the EFT building blocks and scalar quantities constructed out of them, such as $\sigma^\mu_\nu \sigma^\nu_\mu$ and $r^\mu_\nu r^\nu_\mu$. We find that
\begin{equation}\label{eq:con_EFT_building}
\begin{aligned}
&\hat{g}^{\tau \tau} = \frac{g^{\tau \tau}}{f_{0}}\;, \quad 
\hat{K} = \frac{1}{\sqrt{f_{0}}}\bigg(K+\frac{3\dot{f}_0\sqrt{-g^{\tau\tau}}}{2 f_{0}}\bigg)\;, \quad 
\hat{\sigma}^{\mu}_{\nu}= \frac{\sigma^{\mu}_{\nu}}{\sqrt{f_0}}\;, \quad
{}^{(3)}\!\hat{R} = \frac{{}^{(3)}\!R}{f_{0}}\;, \quad  
\hat{r}^{\mu}_{\nu} =  \frac{r^{\mu}_{\nu}}{f_{0}}\;, \\
&\hat{\sigma}^{2}= \frac{\sigma^{2}}{f_{0}}\;, \quad
\hat{\sigma}\hat{r} = \frac{\sigma r}{f_0^{3/2}}\;, \quad 
\hat{r}^2 = \frac{r^2}{f_0^{2}}\;,
\end{aligned}
\end{equation}
where we have defined $\sigma^{2} \equiv \sigma^{\mu}_{\nu} \sigma^{\nu}_{\mu}$, $\sigma r \equiv \sigma^{\mu}_{\nu} r^{\nu}_{\mu} $, and $r^{2} \equiv r^{\mu}_{\nu} r^{\nu}_{\mu}$.
For the rest of the paper, a dot denotes the derivative with respect to $\tau$.
(Namely, $\dot{f}_0=f_{0\Phi}\dot{\Phi}$ with $f_{0\Phi}\equiv {\rm d}f_0/{\rm d}\Phi$.)
Note that $\sqrt{-\hat{g}}=\sqrt{-g}\,f_0^2$.

Let us now study the transformation law of the EFT coefficients.
In doing so, we start from the EFT action in the hatted frame, i.e.,
\begin{align}
\label{EFT_action_hat}
    S= & \int \de^{4}x\sqrt{-\hat{g}}\bigg[\frac{M_{\star}^{2}}{2}\hat{f}(x) \hat{R} - \hat{\Lambda}(x) - \hat{c}(x) \hat{g}^{\tau \tau} - \hat{\beta}(x) \hat{K} - \hat{\alpha}(x)^{\mu}_{\nu} \hat{\sigma}^{\nu}_{\mu}- \hat{\gamma}(x)^{\mu}_{\nu} \hat{r}^{\nu}_{\mu}+\cdots\bigg]\;,
\end{align}
where we have shown only the tadpole terms, and substitute the transformation law~\eqref{eq:con_EFT_building} of the EFT building blocks into it.
Then, the resultant action has the form~\eqref{EFT_action} with each coefficient written in terms of the hatted ones as well as the conformal factor~$f_0$, from which the relation between the unhatted and hatted EFT coefficients can be read off.
Written explicitly, for the tadpole coefficients, we have 
\begin{align}
\begin{split}
&\hat{f}=\frac{f}{f_0}\;, \quad
\hat{\Lambda}=\frac{\Lambda}{f_0^2}+\frac{(2M_\star^2f\bar{K}-3\beta)\dot{f}_0\sqrt{-\bar{g}^{\tau\tau}}}{4f_0^3}-\frac{3M_\star^2f\dot{f}_0^2\bar{g}^{\tau\tau}}{4f_0^4}\;, \\
&\hat{c}=\frac{c}{f_0}+\frac{(2M_\star^2f\bar{K}+3\beta)\dot{f}_0}{4f_0^2\sqrt{-\bar{g}^{\tau\tau}}}\;, \quad
\hat{\alpha}^\mu_\nu=\frac{\alpha^\mu_\nu}{f_0^{3/2}}\;, \quad
\hat{\beta}=\frac{\beta}{f_0^{3/2}}-\frac{M_\star^2f\dot{f}_0\sqrt{-\bar{g}^{\tau\tau}}}{f_0^{5/2}}\;, \quad
\hat{\gamma}^\mu_\nu=\frac{\gamma^\mu_\nu}{f_0}\;.
\end{split}\label{eq:con_tadpole_eft}
\end{align}
Also, for those in front of the terms quadratic in perturbations, we obtain
\begin{align}
&\hat{m}_2^4=m_2^4-\frac{(2M_\star^2f\bar{K}+3\beta+6\bar{g}^{\tau\tau}M_1^3+2\bar{g}^{\tau\tau}\bar{h}^\mu_\nu\lambda^{\nu}_{1\mu})\dot{f}_0}{8f_0(-\bar{g}^{\tau\tau})^{3/2}}+\frac{3(2M_\star^2f-3M_2^2-M_3^2-\bar{h}^\mu_\nu\lambda^{\nu}_{3\mu})\dot{f}_0^2}{16f_0^2\bar{g}^{\tau\tau}}\;, \nonumber \\
&\hat{M}_1^3=\frac{M_1^3}{\sqrt{f_0}}-\frac{(4M_\star^2f-6M_2^2-2M_3^2-\bar{h}^\mu_\nu\lambda^{\nu}_{3\mu})\dot{f}_0}{4f_0^{3/2}\sqrt{-\bar{g}^{\tau\tau}}}\;, \quad
\hat{M}_2^2=\frac{M_2^2}{f_0}\;, \quad
\hat{M}_3^2=\frac{M_3^2}{f_0}\;, \quad
\hat{M}_4=\frac{M_4}{\sqrt{f_0}}\;, \nonumber \\
&\hat{M}_5=\frac{M_5}{\sqrt{f_0}}\;, \quad
\hat{M}_6=\frac{M_6}{\sqrt{f_0}}\;, \quad
\hat{\mu}_1^2=\mu_1^2+\frac{(3M_4+M_5+\bar{h}^\mu_\nu\lambda^{\nu}_{5\mu})\dot{f}_0}{4f_0\sqrt{-\bar{g}^{\tau\tau}}}\;, \quad
\hat{\mu}_2=\mu_2\;, \quad
\hat{\mu}_3=\mu_3\;, \nonumber \\
&\hat{\lambda}^{\nu}_{1\mu}=\frac{\lambda^{\nu}_{1\mu}}{\sqrt{f_0}}+\frac{(2M_6\bar{\sigma}^\nu_\mu+3\lambda^{\nu}_{3\mu})\dot{f}_0}{4f_0^{3/2}\sqrt{-\bar{g}^{\tau\tau}}}\;, \quad
\hat{\lambda}^{\nu}_{2\mu}=\lambda^{\nu}_{2\mu}+\frac{3\lambda^{\nu}_{4\mu}\dot{f}_0}{4f_0\sqrt{-\bar{g}^{\tau\tau}}}\;, \quad
\hat{\lambda}^{\nu}_{3\mu}=\frac{\lambda^{\nu}_{3\mu}}{f_0}\;, \quad
\hat{\lambda}^{\nu}_{4\mu}=\frac{\lambda^{\nu}_{4\mu}}{\sqrt{f_0}}\;, \nonumber \\
&\hat{\lambda}^{\nu}_{5\mu}=\frac{\lambda^{\nu}_{5\mu}}{\sqrt{f_0}}\;, \quad
\hat{\lambda}^{\nu}_{6\mu}=\lambda^{\nu}_{6\mu}\;.
\label{eq:con_second_eft}
\end{align}
Note that, in Eqs.~(\ref{eq:con_tadpole_eft}) and (\ref{eq:con_second_eft}), the function~$f_0$ as well as its derivative are evaluated on the background. Also, it should be emphasized that the transformations~(\ref{eq:con_tadpole_eft}) and (\ref{eq:con_second_eft}) can be applied to an arbitrary background metric.

Some comments are in order.
First, it is straightforward to verify the consistency between the above transformations of the EFT coefficients and those of Horndeski functions presented in Appendix~\ref{HorndeskiConformal}, with the dictionary obtained in \cite{Mukohyama:2022enj}.
Second, the tadpole cancellation conditions as well as the consistency relations should be preserved under an invertible conformal/disformal transformation.
Indeed, if they are satisfied in the hatted frame, then one can confirm that they are automatically satisfied in the unhatted frame, and vice versa.
Third, the above transformation laws of the EFT coefficients are invariant under the following replacements:
    \begin{align}
    (f_0;\bar{g}^{\tau\tau},\bar{K},\cdots;f,\Lambda,c,\cdots;\hat{f},\hat{\Lambda},\hat{c},\cdots)
    \to (1/f_0;\hat{\bar{g}}^{\tau\tau},\hat{\bar{K}},\cdots;\hat{f},\hat{\Lambda},\hat{c},\cdots;f,\Lambda,c,\cdots)\;,
    \end{align}
which manifests that the hatted and unhatted frames play symmetric role.
Here, $f_0$ is replaced by $1/f_0$ since the inverse transformation of Eq.~\eqref{tonlyconf} is given by $g_{\mu\nu}=(1/f_0)\hat{g}_{\mu\nu}$.

\subsection{\texorpdfstring{$X$-independent disformal transformation}{X-independent disformal transformation}}
\label{Section_Disformal_t_only}

In this Subsection, we consider a pure disformal transformation of the form
\begin{equation}
    \hat{g}_{\mu \nu}= 
    g_{\mu \nu}+f_{1}(\Phi)\partial_{\mu}\Phi \partial_{\nu}\Phi \;,
    \label{tonlydisf}
\end{equation}
where $f_1$ is an arbitrary function of $\Phi$ (i.e., function of $\tau$ in the unitary gauge).
Note that any $X$-independent conformal/disformal transformation can be expressed as a functional composition of the pure conformal transformation~\eqref{tonlyconf} and the pure disformal transformation~\eqref{tonlydisf}.
Indeed, for $\hat{g}_{\mu\nu}=f_0(\Phi)\tilde{g}_{\mu\nu}$ and $\tilde{g}_{\mu\nu}=g_{\mu\nu}+f_1(\Phi)\partial_{\mu}\Phi\partial_{\nu}\Phi$, we have
    \begin{align}
    \hat{g}_{\mu\nu}=f_0(\Phi)g_{\mu\nu}+\tilde{f}_1(\Phi)\partial_{\mu}\Phi\partial_{\nu}\Phi\;,
    \end{align}
with $\tilde{f}_1\equiv f_0f_1$.
Therefore, if one would like to know how the EFT is transformed under a conformal/disformal transformation, it is sufficient to study the pure conformal and pure disformal transformations.

As in the previous Subsection, let us first express the disformal transformations of the EFT building blocks as well as scalar quantities constructed out of them.
A straightforward computation yields
\begin{equation}
\begin{aligned}
&\hat{g}^{\tau \tau} = 
\frac{g^{\tau\tau}}{Y}\;, \quad
\hat{K}= \frac{K}{\sqrt{Y}}\;, \quad
\hat{\sigma}^{\mu}_{\nu}= \frac{\sigma^{\mu}_{\nu}}{\sqrt{Y}}\;, \quad
{}^{(3)}\!\hat{R} =  \frac{{}^{(3)}\!R}{f_{0}}\;, \quad
\hat{r}^{\mu}_{\nu} =  \frac{r^{\mu}_{\nu}}{f_{0}}\;, \\
&\hat{\sigma}^{2} = \frac{\sigma^{2}}{Y}\;, \quad
\hat{\sigma}\hat{r} = \frac{\sigma r}{{f_{0} \sqrt{Y}}}\;, \quad
\hat{r}^2= \frac{r^2}{{f_{0}}^{2}} \;,
\end{aligned}
\end{equation}
where $Y \equiv 1+X f_1$.
Note also that $\sqrt{-\hat{g}}=\sqrt{-g}\sqrt{Y}$.
Following the same steps as described in Subsection~\ref{subsectiontonlyconf}, we straightforwardly obtain the catalog for the disformal transformation of EFT parameters. 
For the coefficients up to the first order in perturbations in the action~(\ref{EFT_action}), we find
\begin{align}
&\hat{f}=\frac{f}{\sqrt{\bar{Y}}}\;, \quad
\hat{\Lambda}=\frac{1+\bar{Y}}{2\sqrt{\bar{Y}}}\Lambda-\frac{f_1\bar{X}}{4\bar{Y}^{3/2}}\bigg\{M_\star^2f\bigg[\bar{Y}{}^{(3)}\!\bar{R}-(2+\bar{Y})\bigg(\bar{\sigma}^2-\frac{2}{3}\bar{K}^2\bigg)\bigg]+2\bar{Y}(c\bar{g}^{\tau\tau}-\gamma^\mu_\nu\bar{r}^\nu_\mu)\bigg\}\;, \nonumber \\
&\hat{c}=\frac{(1+\bar{Y})\sqrt{\bar{Y}}}{2}c+\frac{f_1\bar{X}\sqrt{\bar{Y}}}{4\bar{g}^{\tau\tau}}\bigg[M_\star^2f\bigg({}^{(3)}\!\bar{R}-\bar{\sigma}^2+\frac{2}{3}\bar{K}^2\bigg)-2(\Lambda+\gamma^\mu_\nu\bar{r}^\nu_\mu)\bigg]\;, \nonumber \\
&\hat{\alpha}^\mu_\nu=\alpha^\mu_\nu-\frac{f_1\bar{X}}{\bar{Y}}M_\star^2f\bar{\sigma}^\mu_\nu\;, \quad
\hat{\beta}=\beta+\frac{2f_1\bar{X}}{3\bar{Y}}M_\star^2f\bar{K}\;, \quad
\hat{\gamma}^\mu_\nu=\frac{\gamma^\mu_\nu}{\sqrt{\bar{Y}}}\;.
\end{align}
For those associated with the second-order terms, we obtain 
{\small
\begin{align}
&\hat{m}_2^4=\bar{Y}^{7/2}m_2^4-\frac{f_1\bar{X}\bar{Y}^{3/2}}{4\bar{g}^{\tau\tau}}\big[(1+3\bar{Y})c-2\bar{Y}(M_1^3\bar{K}+\lambda^\nu_{1\mu}\bar{K}^\mu_\nu)\big] \nonumber \\
&\qquad -\frac{f_1^2\bar{X}^2\bar{Y}^{3/2}}{8(\bar{g}^{\tau\tau})^2}\big[M_\star^2f({}^{(3)}\!\bar{R}-3\bar{\sigma}^2+2\bar{K}^2)-2(\Lambda+\gamma^\mu_\nu\bar{r}^\nu_\mu+M_2^2\bar{K}^2+M_3^2\bar{K}^\mu_\nu\bar{K}^\nu_\mu+M_6\bar{\sigma}^\mu_\nu\bar{K}^\nu_\alpha\bar{K}^\alpha_\mu+\lambda^\nu_{3\mu}\bar{K}\bar{K}^\mu_\nu)\big]\;, \nonumber \\
&\hat{M}_1^3=\bar{Y}^2M_1^3-\frac{f_1\bar{X}\bar{Y}}{2\bar{g}^{\tau\tau}}\big[2\bar{K}(M_\star^2f-M_2^2)-\lambda^\nu_{3\mu}\bar{K}^\mu_\nu\big]\;, \quad
\hat{M}_2^2=\sqrt{\bar{Y}}\,M_2^2-\frac{f_1\bar{X}}{\sqrt{\bar{Y}}}M_\star^2f\;, \nonumber \\
&\hat{M}_3^2=\sqrt{\bar{Y}}\,M_3^2+\frac{f_1\bar{X}}{\sqrt{\bar{Y}}}M_\star^2f\;, \quad
\hat{M}_4=M_4\;, \quad
\hat{M}_5=M_5\;, \quad
\hat{M}_6=\bar{Y}M_6\;, \nonumber \\
&\hat{\mu}_1^2=\bar{Y}^{3/2}\mu_1^2-\frac{f_1\bar{X}\sqrt{\bar{Y}}}{6\bar{g}^{\tau\tau}}\big(3M_\star^2f+2\bar{h}^\nu_\mu\gamma^\mu_\nu-3M_4\bar{K}-3\lambda^\nu_{5\mu}\bar{K}^\mu_\nu\big)\;, \quad
\hat{\mu}_2=\frac{\mu_2}{\sqrt{\bar{Y}}}\;, \quad
\hat{\mu}_3=\frac{\mu_3}{\sqrt{\bar{Y}}}\;, \nonumber \\
&\hat{\lambda}^\nu_{1\mu}=\bar{Y}^2\lambda^\nu_{1\mu}+\frac{f_1\bar{X}\bar{Y}}{2\bar{g}^{\tau\tau}}\big[2(M_\star^2f+M_3^2)\bar{K}^\nu_\mu+2M_6\bar{\sigma}^\nu_\alpha\bar{K}^\alpha_\mu+\bar{K}\lambda^\nu_{3\mu}\big]\;, \quad
\hat{\lambda}^\nu_{3\mu}=\sqrt{\bar{Y}}\,\lambda^\nu_{3\mu}\;, \quad
\hat{\lambda}^\nu_{4\mu}=\lambda^\nu_{4\mu}\;, \nonumber \\
&\hat{\lambda}^\nu_{2\mu}=\bar{Y}^{3/2}\lambda^\nu_{2\mu}+\frac{f_1\bar{X}\sqrt{\bar{Y}}}{2\bar{g}^{\tau\tau}}\big(2\gamma^\nu_\mu+M_5\bar{K}^\nu_\mu+\bar{K}\lambda^\nu_{4\mu}\big)\;, \quad
\hat{\lambda}^\nu_{5\mu}=\lambda^\nu_{5\mu}\;, \quad
\hat{\lambda}^\nu_{6\mu}=\frac{\lambda^\nu_{6\mu}}{\sqrt{\bar{Y}}}\;.
\end{align}
}

We emphasize here again that the transformations obtained in this Section can be applied to an arbitrary background metric.
Note also that, as in the case of pure conformal transformation discussed in the previous Subsection, the hatted and unhatted frames play symmetric role.
As a result, the above transformation laws are invariant under the following replacements:    \begin{align}
    (f_1;\bar{g}^{\tau\tau},\bar{K},\cdots;f,\Lambda,c,\cdots;\hat{f},\hat{\Lambda},\hat{c},\cdots)
    \to (-f_1;\hat{\bar{g}}^{\tau\tau},\hat{\bar{K}},\cdots;\hat{f},\hat{\Lambda},\hat{c},\cdots;f,\Lambda,c,\cdots)\;.
    \end{align}
Here, $f_1$ is replaced by $-f_1$ since the inverse transformation of Eq.~\eqref{tonlydisf} is given by $g_{\mu\nu}=\hat{g}_{\mu\nu}-f_1\partial_\mu\Phi\partial_\nu\Phi$.

\section{\texorpdfstring{Odd-parity black hole perturbations with $M_6$ operator}{Odd-parity black hole perturbations with M6 operator}}
\label{Section_constrain_c_gw}

In this Section, we study how the EFT operator~$M_6(x)\bar{\sigma}^\mu_\nu \delta K^\nu_\alpha \delta K^\alpha_\mu$, which was introduced in Eq.~\eqref{EFT_action}, affects the dynamics of BH perturbations.
[Recall that the parameter~$\tilde{M}_6^2(x)^\mu_\nu$ therein is now chosen to be $\tilde{M}_6^2(x)^\mu_\nu=M_6(x)\bar{\sigma}^\mu_\nu$ for demonstration purposes.]
In particular, we study odd-parity perturbations about a static and spherically symmetric BH.
As mentioned earlier, since $\bar{\sigma}^\mu_\nu=0$ on a homogeneous and isotropic background, the $M_6$ operator can be used to distinguish phenomena happening at cosmological scales and those observed in the BH regime.\footnote{Similarly, other EFT operators with coefficients involving the background values of $\sigma^\mu_\nu$ and $r^\mu_\nu$ could lead to potential distinctions of phenomena at BH scales and those at cosmological scales. In the present paper, we focus on the $M_6$ operator as it is the leading-order operator in derivatives among such operators.}
It should also be noted that the $M_6$ operator arises, e.g., from the term~$\nabla_\mu \nabla_\nu \Phi \nabla^\nu \nabla^\lambda \Phi \nabla_\lambda \nabla^\mu \Phi$ in cubic higher-order scalar-tensor theories.

Let us start with a static and spherically symmetric background given by
\begin{equation}\label{eq:bg_metric_sp}
    {\rm d}s^2=-A(r) \de t^2 + \frac{\de r^2}{B(r)} + r^2 \left(\de {\theta}^2 + {\sin}^2\theta\,\de {\phi}^2\right),
\end{equation}
where $A(r)$ and $B(r)$ are functions of the areal radius~$r$. The metric above
can be brought to the so-called Lema\^{i}tre coordinates:
\begin{align}\label{eq:bg_metric_Le}
     {\rm d}s^2=-\de \tau^2 + (1-A) \de \rho^2 + r^2 \left(\de {\theta}^2 + {\sin}^2\theta\,\de {\phi}^2\right),
\end{align}
where we have defined $\tau$ and $\rho$ so that
\begin{align}
    \de \tau = \de t + \sqrt{\frac{1-A}{AB}}~\de r \;, \qquad \de \rho = \de t + \frac{\de r}{\sqrt{AB(1-A)}} \;.
\end{align}
We then see that the coordinate~$r$ is a function of $\rho -\tau$, satisfying
\begin{align}
    \partial_\rho r = -\dot{r} = \sqrt{\frac{B(1-A)}{A}} \;, \label{drdrho-drdtau}
\end{align}
where a dot denotes the derivative with respect to $\tau$.
Note that, if we assume that the Lema\^{i}tre coordinate~$\tau$ is compatible with the time coordinate in the unitary gauge (i.e., such that $\dot{\bar{\Phi}} = const.$), then $\bar{X}$ is a constant.

With the setup above, we consider the dynamics of linear odd-parity perturbations based on the following EFT action:
\begin{align}\label{eq:EFT_action}
	S_{\rm odd} = \int \de^4x \sqrt{-g} \bigg[&\frac{M_\star^2}{2}R - \Lambda(r) - c(r)g^{\tau\tau} -\tilde{\beta}(r) K - \alpha(r)\bar{K}^{\mu}_\nu K^\nu_{\mu}
    -\zeta(r)\bar{n}^\mu\partial_\mu g^{\tau\tau} \nonumber \\
	&+ \frac{1}{2} M_6(r) \bar{\sigma}^\mu_\nu \delta K^\nu_\alpha \delta K^\alpha_\mu\bigg] \;,
\end{align}
where we have assumed that $\alpha^\mu_\nu$ in Eq.~\eqref{EFT_action} has the form~$\alpha^\mu_\nu=\alpha\bar{\sigma}^\mu_\nu$ and defined $\tilde{\beta} \equiv \beta - \alpha \bar{K}/3$.
Also, we have omitted the terms~$\gamma^\mu_\nu r^\nu_\mu$, $M_{5}\delta K^{\mu}_{\nu} \delta {}^{(3)}\!R^{\nu}_{\mu}$, and $\mu_{3}\delta {}^{(3)}\!R^{\mu}_{\nu} \delta {}^{(3)}\!R^{\nu}_{\mu}$ for simplicity.
Here, we assume that the action respects the shift symmetry of the scalar field~$\Phi$.
Together with the formula~\eqref{drdrho-drdtau} and the fact that each component of the background metric~\eqref{eq:bg_metric_Le} is a function only of $r$, the EFT parameters should be functions of $r$ only.
We recall that the EFT coefficient in front of $R$ [called $f$ in Eq.~\eqref{EFT_action}] is now assumed to be a function only of $\bar{g}^{\tau\tau}$, which is a constant in the present setup.\footnote{If, on the other hand, $\bar{g}^{\tau\tau}$ is not a constant and/or the coefficient in front of $R$ depends on other quantities such as the extrinsic curvature and the 3d Ricci scalar, one then has to perform conformal/disformal transformations to translate the constraints obtained in the Jordan frame into those in the almost Einstein frame.}
In other words, with all the assumptions mentioned above, we are automatically in an almost Einstein frame.
Therefore, we have set $f=1$ in Eq.~\eqref{eq:EFT_action} without loss of generality.
Notice that we have omitted the operator~$ M_3^2(r) \delta K^\mu_\nu \delta K^\nu_\mu$ as its effect on the odd-mode dynamics was investigated in detail in the previous work~\cite{Mukohyama:2022skk}.

Let us now briefly discuss the background equations.
(See \cite{Mukohyama:2022enj,Mukohyama:2022skk} for a complete discussion.)
As usual, the terms in the first line of Eq.~\eqref{eq:EFT_action} (i.e., the tadpole terms) contribute to the background dynamics.
For a given background metric, the background equations provide relations among the EFT parameters~$\Lambda$, $c$, $\tilde{\beta}$, $\alpha$, and $\zeta$.
In particular, in the situation where $A(r) = B(r)$, the functional form of $\alpha(r)$ is fixed by the following equation:
\begin{align}\label{eq:alpha_bg}
    \alpha(r) = M_\star^2 + \frac{3\lambda}{r(2 -2A + rA')} \;,
\end{align}
with $\lambda$ being an integration constant (of mass dimension) and a prime denoting the derivative with respect to $r$. 

Let us proceed to perturbations around the background~$\bar{g}_{\mu\nu}$ and $\bar{\Phi}$. 
In the odd-parity sector, we introduce the perturbations~$\delta g_{\mu\nu} = g_{\mu\nu} - \bar{g}_{\mu\nu}$ as
\begin{equation}
\begin{aligned}
    \delta g_{\tau\tau} &= \delta g_{\tau \rho} = \delta g_{\rho\rho} = 0 \;, \\
    \delta g_{\tau a} &= \sum_{\ell,m} r^2 h_{0,\ell m}(\tau, \rho) E_a^{\ b} \bar{\nabla}_b Y_{\ell m}(\theta, \phi) \;, \\
    \delta g_{\rho a} &= \sum_{\ell,m} r^2 h_{1,\ell m}(\tau, \rho) E_a^{\ b} \bar{\nabla}_b Y_{\ell m}(\theta, \phi) \;, \\
    \delta g_{a b} &= \sum_{\ell,m} r^2 h_{2,\ell m}(\tau, \rho) E_{(a|}^{\ c} \bar{\nabla}_c \bar{\nabla}_{|b)} Y_{\ell m}(\theta, \phi) \;,
\end{aligned}
\end{equation}
where $a,b,\cdots\in\{\theta, \phi\}$ and we have decomposed perturbations in terms of spherical harmonics~$Y_{\ell m}(\theta,\phi)$.
Here, $E_{ab}$ denotes the completely anti-symmetric rank-2 tensor defined on a 2-sphere, and $\bar{\nabla}_a$ is the covariant derivative associated with the unit 2-sphere metric. 
Note that the perturbation of the scalar field belongs to the even-parity sector, and hence we do not take it into account.
In the present paper, we only consider modes with $\ell\ge 2$ where the odd-parity perturbations have one dynamical degree of freedom.\footnote{The dipole ($\ell=1$) perturbations are non-dynamical and correspond to a slow rotation of the BH. See \cite{Mukohyama:2022skk} for a detailed analysis.}
It is important to note that the perturbation~$h_2$ can be removed by a complete gauge fixing, which can be performed at the Lagrangian level~\cite{Motohashi:2016prk}. 
Therefore, we are left with two variables~$h_0$ and $h_1$.
Practically, one can set $m = 0$ without loss of generality thanks to the spherical symmetry of the background, so that the spherical harmonics simply becomes the Legendre polynomials~$P_\ell(\cos\theta)$ up to an overall factor.
Below, the variables~$h_0$ and $h_1$ denote the coefficients of $P_{\ell}(\cos\theta)$.

Following the standard procedure described, e.g., in \cite{Mukohyama:2022skk,Mukohyama:2023xyf}, the quadratic Lagrangian from the action~(\ref{eq:EFT_action}) for odd-parity perturbations~$h_0$ and $h_1$ can be written in the following form:
\begin{align}\label{L2_odd_Ein}
	\frac{2\ell + 1}{2 \pi j^2}\mathcal{L}_2 = p_1 h_0^2 + p_2 h_1^2 + p_3[(\dot{h}_1 - \partial_\rho h_0)^2+2p_4h_1\partial_\rho h_0] \;,
\end{align}
where $j^2 \equiv \ell (\ell + 1)$.
Among the four coefficients, $p_4$ has the form
    \begin{align}
    p_4=\frac{3\bar{\sigma}^\rho_\rho(4\alpha+M_6\bar{\sigma}^\rho_\rho)}{2(4M_\star^2+M_6\bar{\sigma}^\rho_\rho)}\;, \qquad
    \bar{\sigma}^\rho_\rho=\frac{2}{3r}\sqrt{\frac{B}{A(1-A)}}\bigg(1 - A + \frac{rA'}{2}\bigg)\;.
    \label{p4}
    \end{align}
As shown in \cite{Mukohyama:2022skk}, when $p_4$ is a non-trivial function of $r$, a slowly rotating BH solution is not available as a solution in the $\ell=1$ sector. 
Also, as we will explain in the next paragraph, when $p_4$ is a non-vanishing constant, the sound speed in the $\rho$-direction tends to diverge at large $r$.
Therefore, we demand $p_4=0$, which allows us to express $M_6$ in terms of $\alpha$ (as well as the background metric functions~$A$ and $B$).
After imposing $p_4=0$, the other three coefficients are simplified as
\begin{equation}\label{eq:p1_p3}
	p_1 = \frac{1}{2}(j^2-2)r^2 \sqrt{1-A}(M_\star^2+2\alpha) \;, \qquad
	p_2 = -(j^2-2)\frac{r^2M_\star^2}{2\sqrt{1-A}}  \;, \qquad
	p_3 = \frac{r^4(M^2_\star-\alpha)}{2\sqrt{1 - A}} \;.
\end{equation}
Following the steps described in \cite{Mukohyama:2022skk,Mukohyama:2023xyf}, we then introduce an auxiliary field~$\chi$ in the quadratic Lagrangian~(\ref{L2_odd_Ein}) as 
\begin{align}\label{eq:L2_odd_intro_chi}
   \frac{2\ell + 1}{2 \pi j^2}\mathcal{L}_2 = p_1 h_0^2 + p_2 h_1^2
	+ p_3[-\chi^2 + 2\chi (\dot{h}_1 - \partial_\rho h_0)] \;. 
\end{align}
Notice that it is straightforward to recover the Lagrangian~(\ref{L2_odd_Ein}) (with $p_4=0$) by integrating out $\chi$.
From Eq.~(\ref{eq:L2_odd_intro_chi}), one can integrate out the variables~$h_0$ and $h_1$, giving
\begin{align}
    h_0 = -\frac{\partial_\rho(p_3 \chi)}{p_1} \;, \qquad h_1 = \frac{(p_3 \chi)^{\boldsymbol{\cdot}}}{p_2} \;.
\end{align}
Then, by plugging the above solutions for $h_0$ and $h_1$ in (\ref{eq:L2_odd_intro_chi}), we obtain the quadratic Lagrangian for $\chi$ as
\begin{align}
\frac{(j^2 - 2)(2\ell +1)}{2\pi j^2} \mathcal{L}_2 = s_1 \dot{\chi}^2 - s_2 (\partial_\rho \chi)^2 - s_3 \chi^2 \;,
\end{align}
where the parameters~$s_i$'s are defined as 
\begin{align}
    s_1 \equiv -\frac{(j^2 -2)p_3^2}{p_2} \;, \qquad s_2 \equiv \frac{(j^2 - 2)p_3^2}{p_1} \;, \qquad s_3 \equiv (j^2 - 2)p_3 \bigg[1 - \bigg(\frac{p_1+p_2}{p_1p_2}\dot{p}_3\bigg)^{\boldsymbol{\cdot}}\bigg] \;.
\end{align}
The variable~$\chi$ now represents the propagating degree of freedom in the odd-parity sector. As usual, we define the speed of GW in the radial and angular directions as
\begin{align}
    c_\rho^2 &= \frac{\bar{g}_{\rho\rho}}{|\bar{g}_{\tau\tau}|} \frac{s_2}{s_1} = \frac{M^2_\star}{M_\star^2 + 2 \alpha} \;, \label{eq:c_rho_M6} \\ 
    c_\theta^2 &= \lim_{\ell \rightarrow \infty} \frac{r^2}{|\bar{g}_{\tau\tau}|} \frac{s_3}{j^2 s_1} = \frac{M^2_\star}{M_\star^2 - \alpha} \;. \label{eq:c_theta_M6}
\end{align}
Note that the absence of ghost and gradient instabilities requires that $s_1$, $c_\rho^2$, and $c_\theta^2$ are positive definite.
Interestingly, from Eqs.~(\ref{eq:c_rho_M6}) and (\ref{eq:c_theta_M6}), we see that the speeds in the radial and the angular directions are different, reflecting the traceless nature of $\bar{\sigma}^\mu_\nu$ involved in the $M_6$ operator. 
In particular, we find that $(c_\rho^{-2} - 1) + 2(c_\theta^{-2} - 1) = 0$, where the factor of two in front of the second term comes from the fact that there are two angular directions.
Following \cite{Mukohyama:2023xyf}, let us introduce
\begin{align}
    \alpha_T^{(\rho)}(r) &\equiv c_\rho^2 - 1 =  - \frac{2\alpha}{M_\star^2 + 2 \alpha}\;, \label{eq:alpha_T_rho} \\ 
    \alpha_T^{(\theta)}(r) &\equiv c_\theta^2 - 1 = \frac{\alpha}{M_\star^2 - \alpha} \label{eq:alpha_T_theta} \;,
\end{align}
which characterize the deviation of GW speed from unity in the radial and the angular directions, respectively.

Let us comment on the case where the coefficient~$p_4$ in Eq.~\eqref{L2_odd_Ein} is non-vanishing.
In this case, a straightforward computation yields
\begin{align}
c_\rho^2&=\bigg(\frac{M_\star^2+2\alpha}{M_\star^2}-\frac{2p_4}{M_\star^2\bar{\sigma}^\rho_\rho}\bigg)^{-1}\bigg[1-\frac{2p_4}{3\bar{\sigma}^\rho_\rho}+\frac{r^2p_4^2(M_\star^2-\alpha)}{j^2-2}\bigg]\;, \\
c_\theta^2&=\frac{M_\star^2}{M_\star^2-\alpha}\bigg(1-\frac{2p_4}{3\bar{\sigma}^\rho_\rho}\bigg)\;.
\end{align}
Provided that $\bar{\sigma}^\rho_\rho\propto r^{-3/2}$ and $\alpha\to const.$ at large $r$, which are the case for an asymptotically Schwarzschild background spacetime, we see that $c_\rho^2$ and/or $c_\theta^2$ diverge at spatial infinity unless $p_4$ decays at least as fast as $r^{-3/2}$.
However, as mentioned earlier, when $p_4$ is a non-trivial function of $r$, the EFT does not accommodate a slowly rotating BH solution in the $\ell=1$ sector~\cite{Mukohyama:2022skk}.
This is the reason why we have set $p_4=0$ in our analysis.

Let us now write down the master equation for the odd-parity perturbations, i.e., the generalized RW equation.
For this purpose, we introduce the tortoise coordinate~$r_*$ by
    \begin{align}
    \frac{\de r}{\de r_*} =\sqrt{\frac{B}{A}}\frac{A+\alpha_T^{(\rho)}}{\sqrt{1+\alpha_T^{(\rho)}}}\equiv F(r)\;. \label{def_F}
    \end{align}
We note that the horizon for the odd mode, $r=r_g$, satisfies the condition~$F(r_g) = 0$.
Then, the generalized RW equation is written in the form
    \begin{align}
    \left[\frac{\partial^2}{\partial r_*^2}-\frac{\partial^2}{\partial \tilde{t}^2} - V_\mathrm{eff}(r)\right]\Psi(\tilde{t},r_*)=0\;,
    \end{align}
where we have defined $\Psi \equiv (s_1 s_2)^{1/4} \chi$ and
    \begin{align}
    \tilde{t} \equiv t+\int \de r\sqrt{\frac{1-A}{AB}}\frac{\alpha_T^{(\rho)}}{A+\alpha_T^{(\rho)}} \;.
    \end{align}
Also, the effective potential is given by    
	\begin{align}\label{eq:eff_potential}
	V_{\rm eff}(r)
	=\sqrt{1+\alpha_T^{(\rho)}}\,F \left\{\sqrt{\frac{A}{B}}\frac{1+\alpha_T^{(\theta)}}{1+\alpha_T^{(\rho)}}\frac{\ell(\ell+1)-2}{r^2}+\frac{r}{\big(1+\alpha_T^{(\rho)}\big)^{3/4}}\left[ F \left(\frac{\big(1+\alpha_T^{(\rho)}\big)^{1/4}}{r}\right)'\,\right]'\right\} \;.
	\end{align}
We see that, when $\alpha_T^{(\rho)} = \alpha_T^{(\theta)}$ (i.e., when $c_\rho^2 = c_\theta^2$), the expression above reduces to the one obtained in \cite{Mukohyama:2023xyf}.
Notice that up to now we have not imposed $A(r) = B(r)$.
It is interesting to emphasize that the $\ell$-dependent term of the potential above is affected by the difference between $c_\rho^2$ and $c_\theta^2$.
In other words, the potential with $c_\rho^2 \neq c_\theta^2$ will have different shapes compared to that with $c_\rho^2 = c_\theta^2$.
Therefore, this would affect observables such as the spectrum of QNMs, the tidal Love numbers and the graybody factors, though a detailed analysis of them is beyond the scope of this paper.
Moreover, these effects are anticipated to appear in the gravitational waveforms detected by observations. 
We leave this investigation to future work.
%Add the implications about $c_T^{(\rho)} \neq c_T^{(\theta)}$ in the $\ell$ dependent term in the effective potential, leading to QNM, Love number, etc.
%One can use the generalized RW equation to determine, for instance, the spectrum of QNMs and the tidal Love numbers, though this is beyond the scope of this paper.

For illustrative purposes, let us consider the Hayward metric~\cite{Hayward:2005gi}, which is known as an example of regular BHs:
\begin{align}\label{eq:hayward}
A(r) = B(r) = 1 - \frac{\mu r^2}{r^3 + \sigma^3} \;.
\end{align}
Here, $\mu\,(> 0)$ and $\sigma$ are parameters of length dimension. Note that $\mu$ corresponds to twice the ADM mass and $\sigma$ corresponds to the regularization scale when it is positive.
Indeed, we see that such a metric asymptotically reduces to the Schwarzschild metric as $r \rightarrow \infty$, whereas it goes to the de Sitter metric as $r \rightarrow 0$.
For $\sigma<0$, the metric has a curvature singularity at some finite $r$, but here we regard $\sigma$ as just a phenomenological parameter characterizing the modification from GR and hence allow for negative values of $\sigma$ as well.
For the Hayward metric~\eqref{eq:hayward}, Eq.~(\ref{eq:alpha_bg}) becomes
\begin{align}\label{eq:tadpole_alpha}
\alpha = M_\star^2 + \frac{\lambda (r^3 + \sigma^3)^2}{\mu r^6} \;,
\end{align}
which fixes the functional form of $\alpha_T^{(\rho)}$ and $\alpha_T^{(\theta)}$ defined in Eqs.~(\ref{eq:alpha_T_rho}) and (\ref{eq:alpha_T_theta}).
By requiring that $\alpha_T^{(\rho)}$ (and hence $\alpha_T^{(\theta)}$) vanishes at spatial infinity so that the LIGO/Virgo bound on the speed of GW on cosmological scales is satisfied, we obtain $\lambda = - M_\star^2 \mu$.\footnote{Precisely speaking, when we apply the LIGO/Virgo bound, we should move to the Jordan frame in principle. However, given that our almost Einstein frame is related to the Jordan frame via conformal transformation, we can directly apply the LIGO/Virgo bound in the almost Einstein frame since the speed of light remains the same as in the Jordan frame. Note also that, if one performs a $\Phi$-dependent conformal (or disformal) transformation on the metric~\eqref{eq:bg_metric_Le}, the resultant metric explicitly depends on $\tau$. For a $\Phi$-independent conformal/disformal transformation of a static and spherically symmetric metric, see \cite{Takahashi:2019oxz,BenAchour:2020wiw}.}
As a result, we have
\begin{align}
   \alpha_T^{(\rho)}(r) = \frac{2 \sigma^3 (2r^3 + \sigma^3)}{r^6 - 4 \sigma^3 r^3 - 2\sigma^6}\;, \qquad  
    \alpha_T^{(\theta)}(r) = -\frac{\sigma^3(2r^3 + \sigma^3)}{(r^3 + \sigma^3)^2} \;. \label{eq:alpha_T_hayward} 
\end{align}
Then, we obtain $F(r)$ through Eq.~\eqref{def_F}, whose single zero defines the odd-mode horizon~$r=r_g$.
This allows us to express $\mu$ as
\begin{align}\label{eq:mu_hayward}
\mu =  \frac{r_g (1 + \eta)}{1 - 4\eta - 2\eta^2} \;,
\end{align}
where we have defined $\eta \equiv \sigma^3/r_g^3$.
Note that, in order for $r_g$ to be positive, we demand that $-1 < \eta < \eta_c$ with $\eta_c\equiv \sqrt{3/2}-1\simeq 0.224$.\footnote{We have $r_g>0$ also for $\eta<-\sqrt{3/2}-1\simeq -2.22$, but we ignore this possibility as we are interested in a small deviation from GR ($\eta=0$). Note also that ghost/gradient instabilities are absent for $-1 < \eta < \eta_c$.}
In terms of the dimensionless radial coordinate~$x\equiv r/r_g$, the functional form of $F(x)$ is given by
\begin{align}\label{eq:F_Hayward}
    F(x)=\frac{x^3}{\sqrt{x^6-4\eta x^3 -2\eta^2}}-\frac{1+\eta}{1-4\eta-2\eta^2}\frac{\sqrt{x^6-4\eta x^3 -2\eta^2}}{x(x^3+\eta)}\;.
\end{align}
By construction, we have $F(x = 1) = 0$.
It is also possible to rewrite Eq.~(\ref{eq:alpha_T_hayward}) in terms of dimensionless variables as
\begin{align}\label{eq:alpha_T_x}
 \alpha_T^{(\rho)}(x) = \frac{2\eta (2 x^3 + \eta)}{x^6 - 4\eta x^3 - 2\eta^2} \;, \qquad  
    \alpha_T^{(\theta)}(x) = -\frac{\eta(2 x^3 + \eta)}{(x^3 + \eta)^2} \;. 
\end{align}
Clearly, when $\eta = 0$, we have $\alpha_T^{(\rho)} = \alpha_T^{(\theta)} = 0$ (i.e., no deviation from GR).
We recall that the EFT coefficient~$M_6$ is related to $\alpha$ via the condition~$p_4=0$, with $p_4$ given in Eq.~\eqref{p4}. In terms of the dimensionless variables defined above, we have
    \begin{align}
    M_6=4M_{\star}^2r_g\eta\sqrt{\frac{1-4\eta+2\eta^2}{1+\eta}}\,\frac{(x^3+\eta)^{3/2}(2x^3+\eta)}{x^9}\;,
    \end{align}
which means that the parameter~$\eta$ controls the effect of $M_6$.

\begin{figure}
\centering
  \includegraphics[width=0.49\textwidth]{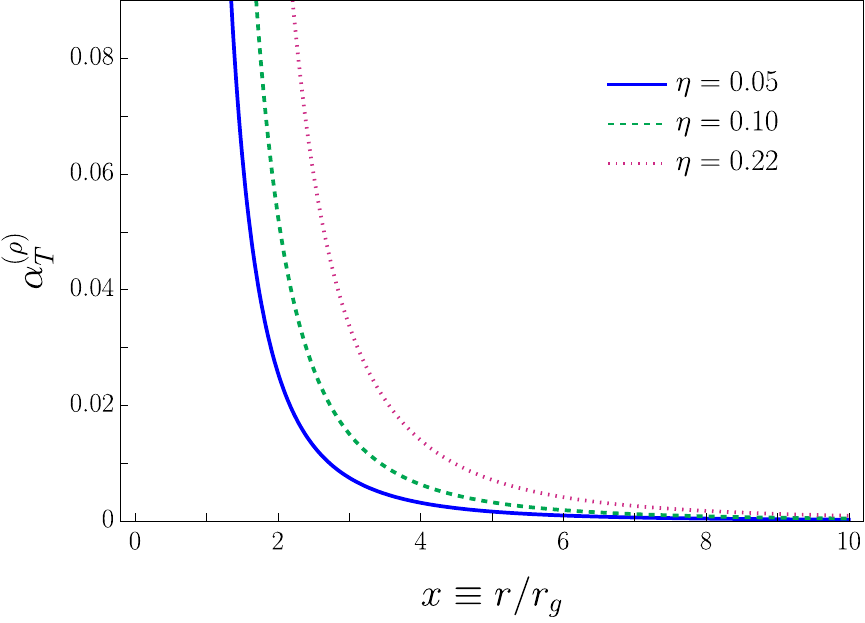} \,
    \includegraphics[width=0.49\textwidth]{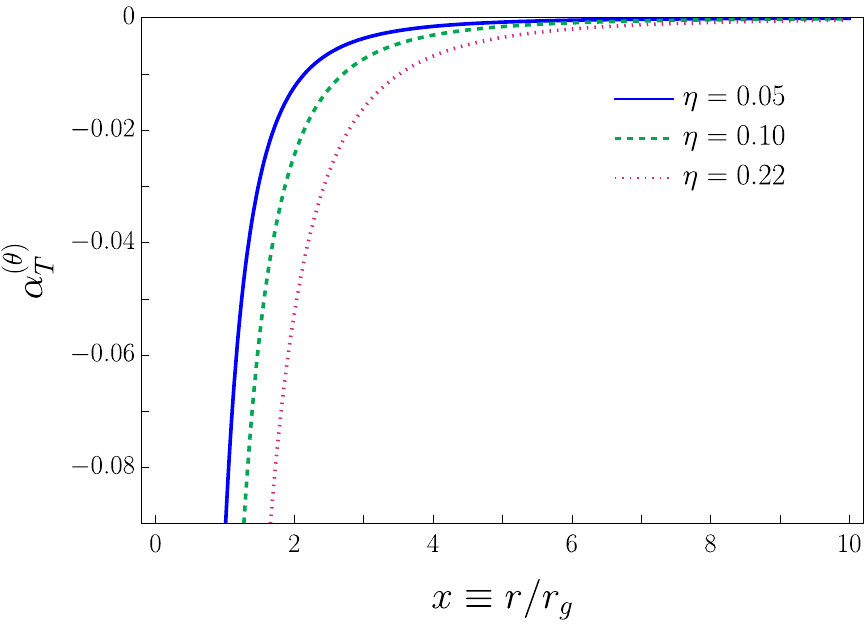}
   \caption{~The parameters~$\alpha_T^{(\rho)}$ (left panel) and $\alpha_T^{(\theta)}$ (right panel) defined in Eq.~(\ref{eq:alpha_T_x}) with various values of $\eta$.
   For both panels, the blue solid, green dashed, and pink dotted curves correspond to $\eta = 0.05$, $0.10$, and $0.22$, respectively. Both $\alpha_T^{(\rho)}$ and $\alpha_T^{(\theta)}$ asymptotically go to zero, by construction.}
  \label{fig:alpha_T} 
\end{figure}

In Figure~\ref{fig:alpha_T}, we plot $\alpha_T^{(\rho)}(x)$ and $\alpha_T^{(\theta)}(x)$ for several values of $\eta$ in the range~$-1 < \eta <\eta_c \simeq 0.224$.
In the plot for both panels, the blue solid, green dashed, and pink dotted curves correspond to $\eta = 0.05$, $0.10$, and $0.22$, respectively.
As mentioned above, both $\alpha_T^{(\rho)}(x)$ and $\alpha_T^{(\theta)}(x)$ go to zero at spatial infinity.
For $0<\eta<\eta_c$, we have $c_\rho^2>1$ (superluminal propagation), while $c_\theta^2<1$ (subluminal propagation).

\begin{figure}
\centering
  \includegraphics[width=0.65\textwidth]{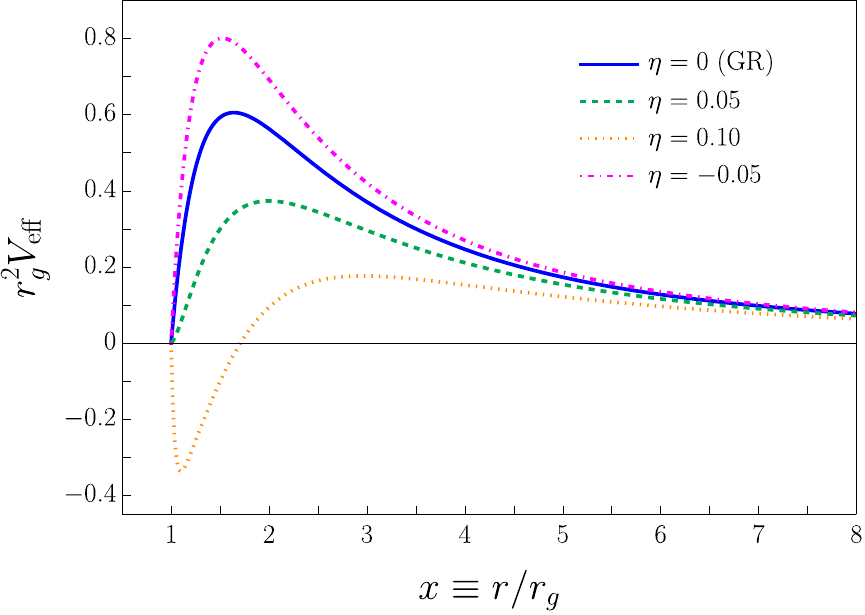} \,
   \caption{~The effective potential for the Hayward metric with several values of $\eta$. The blue solid curve refers to the standard RW potential with $\eta = 0$, whereas the green dashed, orange dotted, and pink dot-dashed curves correspond to $\eta = 0.05$, $0.10$, and $-0.05$.
   The potential goes to zero as $x \rightarrow 1$ ($r \rightarrow r_g$), while it falls off to zero as $r \rightarrow \infty$.}
  \label{fig:V_eff_Hayward} 
\end{figure}

Let us also study the effective potential~(\ref{eq:eff_potential}) on the Hayward background.
Using the formulae~(\ref{eq:F_Hayward}) and (\ref{eq:alpha_T_x}) in Eq.~(\ref{eq:eff_potential}), we obtain
{\small
    \begin{align}
    r_g^2V_{\rm eff}(x)
    =F(x)\frac{\sqrt{x^6-4\eta x^3-2\eta^2}}{x^3+\eta}
    \bigg[&\frac{\ell(\ell+1)x}{x^3+\eta}
    +\frac{3\eta x(12x^{15}-14\eta x^{12}+22\eta^2 x^9+51\eta^3 x^6+28\eta^4 x^3+7\eta^5)}{(x^3+\eta)(x^6-4\eta x^3-2\eta^2)^3} \nonumber \\
    &-\frac{3(1+\eta)}{1-4\eta-2\eta^2}\frac{x^{15}-3\eta x^{12}+11\eta^2 x^9-2\eta^3 x^6-9\eta^4 x^3-\eta^5}{x^3(x^3+\eta)(x^6-4\eta x^3-2\eta^2)^2}\bigg]\;.
    \label{Veff}
    \end{align}
}%
Note that $V_{\rm eff}(x=1)=0$ and the effective potential reduces to the standard RW potential for the Schwarzschild solution in GR when $\eta=0$.
We plot the above potential for $\ell = 2$ in Figure~\ref{fig:V_eff_Hayward} with various values of $\eta$. 
In the plot, the blue solid curve represents the standard RW potential ($\eta = 0$), while the green dashed, orange dotted, and pink dot-dashed curves refer to $\eta = 0.05$, $0.10$, and $-0.05$, respectively.
As explained above, the potential drops to zero as $x \rightarrow 1$ ($r \rightarrow r_g$), while it falls off to zero in the asymptotic regime ($r \rightarrow \infty$).
We see that the modifications to the potentials happen only around the horizon. 
More interestingly, when $\eta$ increases towards the critical value ($\eta=\eta_c \simeq 0.224$), the potential starts developing a dip in the negative region ($V_{\rm eff} < 0$). 
We expect that this modification of the potential might leave interesting features on, for example, the spectrum of QNMs or the tidal Love numbers, which can be potentially observed in the future GW observations.
We leave this investigation to future work.

%%%%%%%%%%%%%%%%%%%%%%%%%%%%%%%%%%%%%%%%%%%%%%%%%%%%%%%%%%%%%%%%%%%%%%%%
\section{Conclusions} \label{sec:conclusions} 
We have studied the conformal/disformal transformation of the coefficients of the effective field theory (EFT) of perturbations on an arbitrary background with a timelike scalar profile. Our main goal in this paper is to provide a connection between the EFT coefficients in the Jordan frame and those in an almost Einstein frame where the coefficient in front of the Einstein-Hilbert term is a constant (or those in any other frame).
By doing so, given the constraints on the parameters of EFT of dark energy which is based in the Jordan frame, one can in principle translate them into the language of EFT of black hole (BH) perturbations.
Another motivation of our study is to prepare a framework of the EFT in the presence of matter fields, where the analysis in the Jordan frame would be more convenient as matter fields follow their geodesics.
For instance, such a framework would be applicable to an extreme mass ratio inspiral of a neutron star into a supermassive BH, which is one of the promising targets for gravitational-wave (GW) astronomy with future space-based observatories.

In Section~\ref{EFT_black_hole_perturbation}, we have reviewed the formulation of the EFT of perturbations on a generic background with a timelike scalar profile, following \cite{Mukohyama:2022enj}. 
We have also included an additional operator in the EFT action, namely $\tilde{M}_6^2(x)^\mu_\nu \delta K^\nu_\alpha \delta K^\alpha_\mu$, with $\tilde{M}_6^2(x)^\mu_\nu$ being a traceless tensor [see Eq.~\eqref{EFT_action}].
Note that such an operator is elusive to the LIGO/Virgo bound on gravitational wave speed as there is no non-trivial traceless tensor on a homogeneous and isotropic background.
For demonstration purposes, we have assumed $\tilde{M}_6^2(x)^\mu_\nu = M_6(r) \bar{\sigma}^\mu_\nu$, with $\bar{\sigma}^\mu_\nu$ denoting the traceless part of the background extrinsic curvature.
In Section~\ref{Section_EFT_catalog}, we have explicitly derived the conformal/disformal transformation of the EFT coefficients in the case where the conformal/disformal factors depend only on the scalar field~$\Phi$. We have confirmed that our result is consistent with the transformation of those in Horndeski theories (see Appendix~\ref{HorndeskiConformal}).
It is straightforward to extend our analysis to the case where both the conformal/disformal factors depend on the scalar kinetic term~$X$ (see Appendix~\ref{Section_general_disformal}).
In Section~\ref{Section_constrain_c_gw}, we have investigated the dynamics of odd-parity perturbations on a static and spherically symmetric BH background, with a particular focus on the effect of the newly-introduced operator~$M_6(r) \bar{\sigma}^{\mu}_{\nu} \delta K^{\nu}_{\alpha} \delta K^{\alpha}_{\mu}$.
We have obtained the quadratic action for the master variable in the odd-parity sector and derived the generalized Regge-Wheeler equation, following the procedure in \cite{Mukohyama:2022skk}.
Interestingly, due to the traceless nature of $\bar{\sigma}^\mu_\nu$, the speed of fluctuation in the radial direction~$c_\rho^2$ is generally different from that in the angular direction~$c_\theta^2$. 
As an example, we have considered the Hayward background, in which the metric goes to the de Sitter metric as $r \rightarrow 0$, while it reduces to the Schwarzschild metric as $r \rightarrow \infty$.
In doing so, we have chosen the tadpole coefficients so that both $c_\rho^2$ and $c_\theta^2$ approach unity at spatial infinity.
On the other hand, due to the $M_6$ operator, $c_\rho^2$ and $c_\theta^2$ deviate from unity in the vicinity of the BH, which is characterized by $\alpha_T^{(\rho)}$ and $\alpha_T^{(\theta)}$ [see Eq.~(\ref{eq:alpha_T_x}) and Figure~\ref{fig:alpha_T}].
Moreover, we have obtained the effective potential for the Hayward background [see Eq.~\eqref{Veff} and Figure~\ref{fig:V_eff_Hayward}].
This allows us to investigate, e.g., the spectrum of quasinormal modes (QNMs) and the tidal Love numbers, which can be used to put a constraint on the operator~$M_6$.

There are several future directions we would like to explore. First, as an immediate application of the conformal/disformal transformation, it would be intriguing to investigate the dynamics of perturbations when matter fields, such as neutron stars, are present. This would shed light on, for example, how the QNM spectrum or the tidal Love number gets affected by the matter source or how the GW waveform changes during the inspiral phase.
Second, it is worth studying in detail how the QNM frequencies deviate from those in GR due to the $M_6$ operator and how the latest GW data can constrain it. 
Finally, it is also interesting to apply the approach developed in this paper to the EFT of vector-tensor theories constructed on an arbitrary background~\cite{Aoki:2023bmz}.
This would enlarge the phenomenology of the EFT when matter fields are present in our system.
We leave all of these issues to future work.

%%%%%%%%%%%%%%%%%%       Acknowledgements          %%%%%%%%%%%%%%%%%%%%
\section*{Acknowledgements}
This work was supported by World Premier International Research Center Initiative (WPI), MEXT, Japan. 
The work of S.~M.~was supported by JSPS KAKENHI Grant No.\ JP24K07017.
The work of K.~T.~was supported by JSPS KAKENHI Grant No.\ JP23K13101.
V.~Y. is supported by grants for development of new faculty staff, Ratchadaphiseksomphot Fund, Chulalongkorn University.

%%%%%%%%%%%%%%%%%%       Appendix          %%%%%%%%%%%%%%%%%%%%

\appendix

\section{Conformal transformation of Horndeski functions}
\label{HorndeskiConformal}

In this Appendix, we consider how the functions~$G_i(\Phi,X)$ of the Horndeski theories are transformed under an $X$-independent conformal transformation~$\hat{g}_{\mu\nu}=f_0(\Phi)g_{\mu\nu}$.
Written explicitly, the Horndeski action is given by
\begin{align}\label{eq:Horndeski}
    S = \int {\rm d}^4x \sqrt{-g}\,\sum_{I = 2}^5\mathcal{L}_I \;,
\end{align}
where 
\begin{equation}
\begin{aligned}
    \mathcal{L}_2 &= G_2(\Phi, X) \;, \\
    \mathcal{L}_3 &= G_3(\Phi, X) \Box \Phi \;, \\
    \mathcal{L}_4 &= G_4(\Phi, X) \tilde{R} - 2 G_{4X}(\Phi, X) [(\Box \Phi)^2 - \nabla_\mu \nabla_\nu \Phi \nabla^\mu \nabla^\nu \Phi] \;, \\
    \mathcal{L}_5 &= G_5(\Phi, X) G_{\mu\nu} \nabla^\nu \nabla^\mu \Phi  \\ 
    & \hspace{4mm} + \frac{1}{3} G_{5X} (\Phi, X) [(\Box\Phi)^3 - 3 \Box\Phi \nabla_\mu \nabla_\nu \Phi \nabla^\mu \nabla^\nu \Phi + 2 \nabla_\mu \nabla_\nu \Phi \nabla^\nu \nabla^\lambda \Phi \nabla_\lambda \nabla^\mu \Phi] \;,
\end{aligned}
\end{equation}
with $G_{iX}\equiv \partial G_i/\partial X$ and $X \equiv \nabla^\mu\Phi \nabla_\mu \Phi$.
Here, $\tilde{R}$ and $G_{\mu\nu}$ denote the Ricci scalar and Einstein tensor in 4d, respectively. It is useful to list the conformal transformation of the following quantities:
\begin{align}
    &\sqrt{-\hat{g}}=\sqrt{-g}\,f_0^2\;, \quad
    \hat{X} =\frac{X}{f_0} \;, \quad
    \hat{\Box} \Phi = \frac{1}{f_0}\bigg(\Box \Phi + \frac{f_{0\Phi}}{f_0} 
    X\bigg) \;, \label{tr_law_first} \\
    &\hat{\nabla}_\mu \hat{\nabla}_\nu \Phi \hat{\nabla}^\mu \hat{\nabla}^\nu \Phi = \frac{1}{f_0^2} \bigg(\nabla_\mu \nabla_\nu \Phi \nabla^\mu \nabla^\nu \Phi - \frac{2 f_{0\Phi}}{f_0} \nabla_\mu \nabla_\nu \Phi \nabla^\mu \Phi \nabla^\nu \Phi + \frac{f_{0\Phi}}{f_0} 
    X \Box \Phi + \frac{f_{0\Phi}^2}{f_0^2} 
    X^2\bigg) \;, \\
    & \hat{\nabla}_\mu \hat{\nabla}_\nu \Phi \hat{\nabla}^\nu \hat{\nabla}^\lambda \Phi \hat{\nabla}_\lambda \hat{\nabla}^\mu \Phi = \frac{1}{f_0^3}\bigg( {\nabla}_\mu {\nabla}_\nu \Phi {\nabla}^\nu {\nabla}^\lambda \Phi {\nabla}_\lambda {\nabla}^\mu \Phi -\frac{3 f_{0\Phi}}{f_0}{\nabla}_\mu {\nabla}_\nu \Phi {\nabla}^\nu {\nabla}^\lambda \Phi {\nabla}_\lambda \Phi {\nabla}^\mu \Phi \nonumber \\
    & \hspace{5cm} + \frac{3 f_{0\Phi}}{2 f_0} X  \nabla_\mu \nabla_\nu \Phi \nabla^\mu \nabla^\nu \Phi 
    + \frac{3 f_{0\Phi}^{2}}{4 f_0^{2}} X^2 \Box \Phi 
    + \frac{f_{0\Phi}^{3}}{4 f_0^{3}} 
    X^3 \bigg)\;,
\end{align}
and
\begin{align}   
    \hat{\tilde{R}}_{\mu\nu} &= \tilde{R}_{\mu\nu} - \frac{f_{0\Phi}}{f_0} \nabla_\mu \nabla_\nu \Phi 
    + \bigg(\frac{3 f_{0\Phi}^2}{2f_0^2} - \frac{f_{0\Phi\Phi}}{f_0}\bigg) \nabla_\mu\Phi \nabla_\nu \Phi 
    -g_{\mu\nu}\bigg(\frac{f_{0\Phi}}{2f_0}\Box\Phi+\frac{f_{0\Phi\Phi}}{2f_0}X\bigg)\;, \\
    \hat{\tilde{R}} &= \frac{1}{f_0}\bigg(\tilde{R} - \frac{3f_{0\Phi}}{f_0} \Box \Phi + \frac{3f^2_{0\Phi}}{2 f_0^2} 
    X - \frac{3 f_{0\Phi\Phi}}{f_0} 
    X\bigg) \;, \label{eq:Ricci_conf} \\
    \hat{G}_{\mu\nu} &= G_{\mu\nu} 
    - \frac{f_{0\Phi}}{f_0} \nabla_\mu \nabla_\nu \Phi 
    + \bigg(\frac{3 f^2_{0\Phi}}{2 f_0^2} - \frac{f_{0\Phi\Phi}}{f_0}\bigg) \nabla_\mu \Phi \nabla_\nu \Phi +g_{\mu\nu}\bigg(\frac{f_{0\Phi}}{f_0}\Box\Phi-\frac{3f^2_{0\Phi}}{4f_0^2}X+\frac{f_{0\Phi\Phi}}{f_0}X\bigg)\;. \label{tr_law_last}
\end{align}
Using the expressions above in the Horndeski action~\eqref{eq:Horndeski},\footnote{Precisely speaking, we first prepare a hatted counterpart of the Horndeski action~\eqref{eq:Horndeski} where a hat is put on each of the metric~$g_{\mu\nu}$ and $G_i$'s, and then substitute the transformation laws~\eqref{tr_law_first}--\eqref{tr_law_last} into it and compare the resultant action with Eq.~\eqref{eq:Horndeski}.} we obtain $\hat{G}_4(\Phi, \hat{X}) = G_4(\Phi, X)/f_0$, $\hat{G}_5(\Phi, \hat{X}) = G_5(\Phi, X)$, and 
\begin{align}
G_2+G_3\Box\Phi&=f_0^2\hat{G}_2+f_0\hat{G}_3\Box\Phi+f_{0\Phi}X\hat{G}_3+\bigg(\frac{3f_{0\Phi}^2}{2f_0^2}-\frac{3f_{0\Phi\Phi}}{f_0}\bigg)XG_4-\frac{3f_{0\Phi}}{2f_0}\bigg(\frac{f_{0\Phi}^2}{f_0^2}-\frac{f_{0\Phi\Phi}}{f_0}\bigg)X^2G_5 \nonumber \\
&\quad -\frac{f_{0\Phi}^3}{2f_0^3}X^3G_{5X}-\bigg[\frac{f_{0\Phi}}{f_0}(3G_4+2XG_{4X})+\bigg(\frac{f_{0\Phi}^2}{4f_0^2}-\frac{f_{0\Phi\Phi}}{f_0}\bigg)XG_5+\frac{f_{0\Phi}^2}{2f_0^2}X^2G_{5X}\bigg]\Box\Phi \nonumber \\
&\quad -\bigg[\frac{4f_{0\Phi}}{f_0}G_{4X}-\bigg(\frac{5f_{0\Phi}^2}{2f_0^2}-\frac{f_{0\Phi\Phi}}{f_0}\bigg)G_5-\frac{2f_{0\Phi}^2}{f_0^2}XG_{5X}\bigg]\nabla^\mu\Phi\nabla^\nu\Phi\nabla_\mu\nabla_\nu\Phi \nonumber \\
&\quad -\frac{f_{0\Phi}}{f_0}G_5\big[R_{\mu\nu}\nabla^\mu\Phi\nabla^\nu\Phi-(\Box\Phi)^2+\nabla_\mu\nabla_\nu\Phi\nabla^\mu\nabla^\nu\Phi\big] \nonumber \\
&\quad +\frac{2f_{0\Phi}}{f_0}G_{5X}\nabla^\mu\Phi\nabla^\nu\Phi\big(\Box\Phi\nabla_\mu\nabla_\nu\Phi-\nabla_\mu\nabla^\lambda\Phi\nabla_\lambda\nabla_\nu\Phi\big)\;. \label{eq:G_2_G_3}
\end{align}
By use of several mathematical identities and integration by parts, one can reexpress the right-hand side in the form of $G_2+G_3\Box\Phi$.
Indeed, we have the following identities:
\begin{align}
&H_X\nabla^\mu\Phi\nabla^\nu\Phi\nabla_\mu\nabla_\nu\Phi=-\frac{1}{2}H_{\Phi}X-\frac{1}{2}H\Box\Phi 
+\nabla_\mu\bigg(\frac{H}{2}\nabla^\mu\Phi\bigg)\;, \label{int_by_parts1} \\
&H_X\big[R_{\mu\nu}\nabla^\mu\Phi\nabla^\nu\Phi-(\Box\Phi)^2+\nabla_\mu\nabla_\nu\Phi\nabla^\mu\nabla^\nu\Phi\big]-2H_{XX}\nabla^\mu\Phi\nabla^\nu\Phi\big(\Box\Phi\nabla_\mu\nabla_\nu\Phi-\nabla_\mu\nabla^\lambda\Phi\nabla_\lambda\nabla_\nu\Phi\big) \nonumber \\
&\qquad =\frac{1}{2}H_{\Phi\Phi}X+\bigg(\frac{1}{2}H_{\Phi}+XH_{\Phi X}\bigg)\Box\Phi 
+\nabla_\mu\bigg(2H_X\nabla_\nu\nabla^{[\mu}\Phi\nabla^{\nu]}\Phi-\frac{H_\Phi}{2}\nabla^\mu\Phi\bigg)\;, \label{int_by_parts2}
\end{align}
with $H$ being an arbitrary function of $\Phi$ and $X$ (see \cite{Gleyzes:2013ooa} for a similar trick).
Note that the last term of each equation is a total derivative.
Equation~\eqref{int_by_parts1} can be applied to the third line of Eq.~\eqref{eq:G_2_G_3}, while Eq.~\eqref{int_by_parts2} can be applied to the fourth and fifth lines of Eq.~\eqref{eq:G_2_G_3}.
After some manipulations, we obtain
\begin{align}
f_0\hat{G}_3&=G_3+\frac{f_{0\Phi}}{f_0}G_4+\frac{2f_{0\Phi}}{f_0}XG_{4X}+\frac{f_{0\Phi}^2}{4f_0^2}XG_5+\frac{f_{0\Phi}}{f_0}XG_{5\Phi}+\frac{f_{0\Phi}^2}{2f_0^2}X^2G_{5X}-I\;, \\
f_0^2\hat{G}_2&=G_2-\frac{f_{0\Phi}}{f_0}XG_3-\bigg(\frac{f_{0\Phi}^2}{2f_0^2}-\frac{f_{0\Phi\Phi}}{f_0}\bigg)XG_4-\frac{2f_{0\Phi}}{f_0}XG_{4\Phi}-\frac{2f_{0\Phi}^2}{f_0^2}X^2G_{4X} \nonumber \\
&\quad -\frac{f_{0\Phi}}{f_0}\bigg(\frac{3f_{0\Phi}^2}{4f_0^2}-\frac{f_{0\Phi\Phi}}{2f_0}\bigg)X^2G_5+\frac{f_{0\Phi}}{f_0}XI-XI_{\Phi}\;,
\end{align}
where we have defined
\begin{align}
I(\Phi,X)\equiv \frac{f_{0\Phi}}{2f_0}\int {\rm d}X\bigg(\frac{f_{0\Phi}}{2f_0}G_{5}-G_{5\Phi}\bigg)\;.
\end{align}
Note that the integral~$I$ can be determined up to a $\Phi$-dependent integration constant, whose contribution to the action is a total derivative.
With the dictionary between the Horndeski functions and the EFT parameters obtained in \cite{Mukohyama:2022enj}, we have confirmed that the above results are consistent with those in Section~\ref{subsectiontonlyconf}.

\section{General conformal/disformal transformation of the EFT}
\label{Section_general_disformal}
In this Appendix, we present the basic ingredients to obtain the transformation of the EFT parameters under the general conformal/disformal transformation~(\ref{disf}), where both $f_0$ and $f_1$ are functions of $\Phi$ and $X$.
As implied in Eq.~\eqref{General_Action}, the EFT Lagrangian is given by a scalar function constructed out of $\tau$, $g^{\tau\tau}$, $K^\mu_\nu$, and ${}^{(3)}\!R^\mu_\nu$, as well as their covariant derivatives.
Among them, $K^\mu_\nu$ and ${}^{(3)}\!R^\mu_\nu$ are further decomposed into the trace and traceless parts as $K^\mu_\nu=\sigma^\mu_\nu+Kh^\mu_\nu/3$ and ${}^{(3)}\!R^\mu_\nu=r^\mu_\nu+{}^{(3)}\!Rh^\mu_\nu/3$, respectively.
The transformation of the EFT building blocks is given by
\begin{align}
\begin{split}
&\hat{g}^{\tau\tau}=\frac{g^{\tau\tau}}{Y}\;, \quad
\hat{K}=\frac{1}{\sqrt{Y}}\bigg(K+\frac{3}{2f_0}\pounds_{n}f_0\bigg)\;, \quad
\hat{\sigma}^\mu_\nu=\frac{\sigma^\mu_\nu}{\sqrt{Y}}\;, \\
&{}^{(3)}\!\hat{R}=\frac{1}{f_0}\bigg({}^{(3)}\!R-\frac{2}{f_0}{\rm D}^{\alpha}{\rm D}_{\alpha}f_0+\frac{3}{2f_0^2}{\rm D}^{\alpha}f_0{\rm D}_{\alpha}f_0\bigg)\;, \\
&\hat{r}^{\mu}_{\nu}=\frac{1}{f_0}\bigg[r^{\mu}_{\nu}-\frac{1}{2f_0}{\rm D}^{\mu}{\rm D}_{\nu}f_0+\frac{3}{4f_0^2}{\rm D}^{\mu}f_0{\rm D}_{\nu}f_0+h^\mu_\nu\bigg(\frac{1}{6f_0}{\rm D}^{\alpha}{\rm D}_{\alpha}f_0-\frac{1}{4f_0^2}{\rm D}^{\alpha}f_0{\rm D}_{\alpha}f_0\bigg)\bigg]\;,
\end{split}\label{EFT_trnsf_general_conf/disf}
\end{align}
with $Y=f_0+Xf_1$.
Note also that $\sqrt{-\hat{g}}=\sqrt{-g}\,f_0^{3/2}\sqrt{Y}$.
Here, we have denoted the Lie derivative along $n^\mu$ and the spatial covariant derivative respectively as $\pounds_{n}$ and ${\rm D}_\mu$, i.e.,
\begin{align}
\pounds_{n}f_0=n^\mu\nabla_\mu f_0\;, \qquad
{\rm D}_\mu f_0=h_\mu^\alpha\nabla_\alpha f_0\;, \qquad
{\rm D}_\mu{\rm D}_\nu f_0=h_\mu^\alpha h_\nu^\beta \nabla_\alpha(h_\beta^\gamma \nabla_\gamma f_0)\;.
\end{align}
Note that $X=\dot{\bar{\Phi}}^2g^{\tau\tau}$ in the unitary gauge, and therefore $f_0$ and $f_1$ (and $Y$ as well) are functions of $\tau$ and $g^{\tau\tau}$.
In particular, $\hat{g}^{\tau\tau}$ is a function only of $\tau$ and $g^{\tau\tau}$.
It should also be noted that the set of EFT building blocks~$\{g^{\tau\tau},K,\sigma^\mu_\nu,{}^{(3)}\!R,r^\mu_\nu\}$ itself is not closed under the general conformal/disformal transformation as the transformation law~\eqref{EFT_trnsf_general_conf/disf} contains derivatives of $g^{\tau\tau}$ through $\pounds_{n}f_0$, ${\rm D}_\mu f_0$, and ${\rm D}_\mu{\rm D}_\nu f_0$ when $f_{0X}\ne 0$.
The minimal set of EFT building blocks that is closed under the general conformal/disformal transformation is obtained by including\footnote{Since $\hat{g}^{\tau\tau}$ is a function only of $\tau$ and $g^{\tau\tau}$, a conformal/disformal transformation of the new building blocks in Eq.~\eqref{new_building_blocks} does not yield unwanted extra quantities.}
\begin{align}
\pounds_{n}g^{\tau\tau}\;, \qquad
{\rm D}_\mu g^{\tau\tau}\;, \qquad
{\rm D}_\mu{\rm D}_\nu g^{\tau\tau}\;. \label{new_building_blocks}
\end{align}

As mentioned in the main text, practically, we perform a perturbative expansion of the EFT action constructed out of these building blocks.
In doing so, it is convenient to choose variables so that the action for perturbations becomes simple.
Actually, when we have defined the perturbation of rank-2 tensors in Eq.~\eqref{eq:pert_def}, we have assumed rank-2 tensors with one upper index and one lower index, which allows us to contract rank-2 tensors with the Kronecker delta.
Otherwise, one has to contract tensors with the projection tensor~$h_{\mu\nu}$, and hence the perturbation of $h_{\mu\nu}$ should also be taken into account.
For this reason, regarding the new EFT building block~${\rm D}_\mu{\rm D}_\nu g^{\tau\tau}$ in Eq.~\eqref{new_building_blocks}, it would be useful to consider ${\rm D}^\mu{\rm D}_\nu g^{\tau\tau}$ for the perturbative analysis.
Regarding ${\rm D}_\mu g^{\tau\tau}$, one can consider its 3d dual~$\epsilon^\mu{}_{\nu\alpha\beta}n^\alpha{\rm D}^\beta g^{\tau\tau}$, with $\epsilon_{\mu\nu\alpha\beta}$ being the Levi-Civita tensor.
(See \cite{Aoki:2023bmz} for a related discussion.)

In our EFT action~\eqref{EFT_action}, we have partially included operators with derivatives acting on $g^{\tau\tau}$, which correspond to the terms with $\zeta$, ${\cal M}_1$, ${\cal M}_2$, and ${\cal M}_3$.
Indeed, these terms were introduced in \cite{Mukohyama:2022skk} to incorporate the class of quadratic higher-order scalar-tensor theories, which is closed under the general conformal/disformal transformation.
Once the set of EFT building blocks is specified, it is straightforward to find the transformation law of the EFT coefficients by following the same procedure as in Section~\ref{Section_EFT_catalog}.

%%%%%%%%%%%%%%%%%%%%%%%%%%%%%%%%%%%%%%%%%%%%%%%%%%
{}
\bibliographystyle{utphys}
\bibliography{bib_v4}

\providecommand{\href}[2]{#2}\begingroup\raggedright\begin{thebibliography}{10}

\bibitem{Koyama:2015vza}
K.~Koyama, ``{Cosmological Tests of Modified Gravity},'' {\em Rept. Prog.
  Phys.} {\bf 79} (2016), no.~4 046902,
  \href{https://arxiv.org/abs/1504.04623}{{\tt 1504.04623}}.

\bibitem{Ferreira:2019xrr}
P.~G. Ferreira, ``{Cosmological Tests of Gravity},'' {\em Ann. Rev. Astron.
  Astrophys.} {\bf 57} (2019) 335--374,
  \href{https://arxiv.org/abs/1902.10503}{{\tt 1902.10503}}.

\bibitem{Arai:2022ilw}
S.~Arai {\em et~al.}, ``{Cosmological gravity probes: connecting recent
  theoretical developments to forthcoming observations},'' {\em PTEP} {\bf
  2023} (2023) 072E01, \href{https://arxiv.org/abs/2212.09094}{{\tt
  2212.09094}}.

\bibitem{Lovelock:1971yv}
D.~Lovelock, ``{The Einstein tensor and its generalizations},'' {\em J. Math.
  Phys.} {\bf 12} (1971) 498--501.

\bibitem{Lovelock:1972vz}
D.~Lovelock, ``{The four-dimensionality of space and the einstein tensor},''
  {\em J. Math. Phys.} {\bf 13} (1972) 874--876.

\bibitem{Brans:1961sx}
C.~Brans and R.~H. Dicke, ``{Mach's principle and a relativistic theory of
  gravitation},'' {\em Phys. Rev.} {\bf 124} (1961) 925--935.

\bibitem{Armendariz-Picon:2000nqq}
C.~Armendariz-Picon, V.~F. Mukhanov, and P.~J. Steinhardt, ``{A Dynamical
  solution to the problem of a small cosmological constant and late time cosmic
  acceleration},'' {\em Phys. Rev. Lett.} {\bf 85} (2000) 4438--4441,
  \href{https://arxiv.org/abs/astro-ph/0004134}{{\tt astro-ph/0004134}}.

\bibitem{Armendariz-Picon:2000ulo}
C.~Armendariz-Picon, V.~F. Mukhanov, and P.~J. Steinhardt, ``{Essentials of k
  essence},'' {\em Phys. Rev. D} {\bf 63} (2001) 103510,
  \href{https://arxiv.org/abs/astro-ph/0006373}{{\tt astro-ph/0006373}}.

\bibitem{Horndeski:1974wa}
G.~W. Horndeski, ``{Second-order scalar-tensor field equations in a
  four-dimensional space},'' {\em Int. J. Theor. Phys.} {\bf 10} (1974)
  363--384.

\bibitem{Deffayet:2011gz}
C.~Deffayet, X.~Gao, D.~A. Steer, and G.~Zahariade, ``{From k-essence to
  generalised Galileons},'' {\em Phys. Rev. D} {\bf 84} (2011) 064039,
  \href{https://arxiv.org/abs/1103.3260}{{\tt 1103.3260}}.

\bibitem{Kobayashi:2011nu}
T.~Kobayashi, M.~Yamaguchi, and J.~Yokoyama, ``{Generalized G-inflation:
  Inflation with the most general second-order field equations},'' {\em Prog.
  Theor. Phys.} {\bf 126} (2011) 511--529,
  \href{https://arxiv.org/abs/1105.5723}{{\tt 1105.5723}}.

\bibitem{Langlois:2015cwa}
D.~Langlois and K.~Noui, ``{Degenerate higher derivative theories beyond
  Horndeski: evading the Ostrogradski instability},'' {\em JCAP} {\bf 02}
  (2016) 034, \href{https://arxiv.org/abs/1510.06930}{{\tt 1510.06930}}.

\bibitem{Motohashi:2016ftl}
H.~Motohashi, K.~Noui, T.~Suyama, M.~Yamaguchi, and D.~Langlois, ``{Healthy
  degenerate theories with higher derivatives},'' {\em JCAP} {\bf 07} (2016)
  033, \href{https://arxiv.org/abs/1603.09355}{{\tt 1603.09355}}.

\bibitem{Gleyzes:2014dya}
J.~Gleyzes, D.~Langlois, F.~Piazza, and F.~Vernizzi, ``{Healthy theories beyond
  Horndeski},'' {\em Phys. Rev. Lett.} {\bf 114} (2015), no.~21 211101,
  \href{https://arxiv.org/abs/1404.6495}{{\tt 1404.6495}}.

\bibitem{Crisostomi:2016czh}
M.~Crisostomi, K.~Koyama, and G.~Tasinato, ``{Extended Scalar-Tensor Theories
  of Gravity},'' {\em JCAP} {\bf 1604} (2016), no.~04 044,
  \href{https://arxiv.org/abs/1602.03119}{{\tt 1602.03119}}.

\bibitem{BenAchour:2016fzp}
J.~Ben~Achour, M.~Crisostomi, K.~Koyama, D.~Langlois, K.~Noui, and G.~Tasinato,
  ``{Degenerate higher order scalar-tensor theories beyond Horndeski up to
  cubic order},'' {\em JHEP} {\bf 12} (2016) 100,
  \href{https://arxiv.org/abs/1608.08135}{{\tt 1608.08135}}.

\bibitem{Langlois:2018dxi}
D.~Langlois, ``{Dark energy and modified gravity in degenerate higher-order
  scalar\textendash{}tensor (DHOST) theories: A review},'' {\em Int. J. Mod.
  Phys. D} {\bf 28} (2019), no.~05 1942006,
  \href{https://arxiv.org/abs/1811.06271}{{\tt 1811.06271}}.

\bibitem{Kobayashi:2019hrl}
T.~Kobayashi, ``{Horndeski theory and beyond: a review},'' {\em Rept. Prog.
  Phys.} {\bf 82} (2019), no.~8 086901,
  \href{https://arxiv.org/abs/1901.07183}{{\tt 1901.07183}}.

\bibitem{DeFelice:2018ewo}
A.~De~Felice, D.~Langlois, S.~Mukohyama, K.~Noui, and A.~Wang, ``{Generalized
  instantaneous modes in higher-order scalar-tensor theories},'' {\em Phys.
  Rev. D} {\bf 98} (2018), no.~8 084024,
  \href{https://arxiv.org/abs/1803.06241}{{\tt 1803.06241}}.

\bibitem{DeFelice:2021hps}
A.~De~Felice, S.~Mukohyama, and K.~Takahashi, ``{Nonlinear definition of the
  shadowy mode in higher-order scalar-tensor theories},'' {\em JCAP} {\bf 12}
  (2021), no.~12 020, \href{https://arxiv.org/abs/2110.03194}{{\tt
  2110.03194}}.

\bibitem{DeFelice:2022xvq}
A.~De~Felice, S.~Mukohyama, and K.~Takahashi, ``{Avoidance of Strong Coupling
  in General Relativity Solutions with a Timelike Scalar Profile in a Class of
  Ghost-Free Scalar-Tensor Theories},'' {\em Phys. Rev. Lett.} {\bf 129}
  (2022), no.~3 031103, \href{https://arxiv.org/abs/2204.02032}{{\tt
  2204.02032}}.

\bibitem{Bekenstein:1992pj}
J.~D. Bekenstein, ``{The Relation between physical and gravitational
  geometry},'' {\em Phys. Rev. D} {\bf 48} (1993) 3641--3647,
  \href{https://arxiv.org/abs/gr-qc/9211017}{{\tt gr-qc/9211017}}.

\bibitem{Bruneton:2007si}
J.-P. Bruneton and G.~Esposito-Farese, ``{Field-theoretical formulations of
  MOND-like gravity},'' {\em Phys. Rev. D} {\bf 76} (2007) 124012,
  \href{https://arxiv.org/abs/0705.4043}{{\tt 0705.4043}}. [Erratum:
  \underline{Phys.~Rev.~D} {\bf 76} (2007) 129902].

\bibitem{Bettoni:2013diz}
D.~Bettoni and S.~Liberati, ``{Disformal invariance of second order
  scalar-tensor theories: Framing the Horndeski action},'' {\em Phys. Rev. D}
  {\bf 88} (2013) 084020, \href{https://arxiv.org/abs/1306.6724}{{\tt
  1306.6724}}.

\bibitem{Zumalacarregui:2013pma}
M.~Zumalac\'arregui and J.~Garc\'\i{}a-Bellido, ``{Transforming gravity: from
  derivative couplings to matter to second-order scalar-tensor theories beyond
  the Horndeski Lagrangian},'' {\em Phys. Rev. D} {\bf 89} (2014) 064046,
  \href{https://arxiv.org/abs/1308.4685}{{\tt 1308.4685}}.

\bibitem{Takahashi:2017pje}
K.~Takahashi and T.~Kobayashi, ``{Extended mimetic gravity: Hamiltonian
  analysis and gradient instabilities},'' {\em JCAP} {\bf 11} (2017) 038,
  \href{https://arxiv.org/abs/1708.02951}{{\tt 1708.02951}}.

\bibitem{Langlois:2018jdg}
D.~Langlois, M.~Mancarella, K.~Noui, and F.~Vernizzi, ``{Mimetic gravity as
  DHOST theories},'' {\em JCAP} {\bf 02} (2019) 036,
  \href{https://arxiv.org/abs/1802.03394}{{\tt 1802.03394}}.

\bibitem{Takahashi:2021ttd}
K.~Takahashi, H.~Motohashi, and M.~Minamitsuji, ``{Invertible disformal
  transformations with higher derivatives},'' {\em Phys. Rev. D} {\bf 105}
  (2022), no.~2 024015, \href{https://arxiv.org/abs/2111.11634}{{\tt
  2111.11634}}.

\bibitem{Takahashi:2023vva}
K.~Takahashi, ``{Invertible disformal transformations with arbitrary
  higher-order derivatives},'' {\em Phys. Rev. D} {\bf 108} (2023), no.~8
  084031, \href{https://arxiv.org/abs/2307.08814}{{\tt 2307.08814}}.

\bibitem{Takahashi:2022mew}
K.~Takahashi, M.~Minamitsuji, and H.~Motohashi, ``{Generalized disformal
  Horndeski theories: Cosmological perturbations and consistent matter
  coupling},'' {\em PTEP} {\bf 2023} (2023), no.~1 013E01,
  \href{https://arxiv.org/abs/2209.02176}{{\tt 2209.02176}}.

\bibitem{Takahashi:2023jro}
K.~Takahashi, M.~Minamitsuji, and H.~Motohashi, ``{Effective description of
  generalized disformal theories},'' {\em JCAP} {\bf 07} (2023) 009,
  \href{https://arxiv.org/abs/2304.08624}{{\tt 2304.08624}}.

\bibitem{Domenech:2023ryc}
G.~Dom\`enech and A.~Ganz, ``{Disformal symmetry in the Universe: mimetic
  gravity and beyond},'' {\em JCAP} {\bf 08} (2023) 046,
  \href{https://arxiv.org/abs/2304.11035}{{\tt 2304.11035}}.

\bibitem{Tahara:2023pyg}
H.~W.~H. Tahara, K.~Takahashi, M.~Minamitsuji, and H.~Motohashi, ``{Exact
  solution for rotating black holes in parity-violating gravity},'' {\em PTEP}
  {\bf 2025} (2024), no.~5 053E02, \href{https://arxiv.org/abs/2312.11899}{{\tt
  2312.11899}}.

\bibitem{Babichev:2024eoh}
E.~Babichev, K.~Izumi, K.~Noui, N.~Tanahashi, and M.~Yamaguchi,
  ``{Generalisation of Conformal-Disformal Transformations of the Metric in
  Scalar-Tensor Theories},'' \href{https://arxiv.org/abs/2405.13126}{{\tt
  2405.13126}}.

\bibitem{Naruko:2022vuh}
A.~Naruko, R.~Saito, N.~Tanahashi, and D.~Yamauchi, ``{Ostrogradsky mode in
  scalar\textendash{}tensor theories with higher-order derivative couplings to
  matter},'' {\em PTEP} {\bf 2023} (2023), no.~5 053E02,
  \href{https://arxiv.org/abs/2209.02252}{{\tt 2209.02252}}.

\bibitem{Takahashi:2022ctx}
K.~Takahashi, R.~Kimura, and H.~Motohashi, ``{Consistency of matter coupling in
  modified gravity},'' {\em Phys. Rev. D} {\bf 107} (2023), no.~4 044018,
  \href{https://arxiv.org/abs/2212.13391}{{\tt 2212.13391}}.

\bibitem{Ikeda:2023ntu}
T.~Ikeda, K.~Takahashi, and T.~Kobayashi, ``{Consistency of higher-derivative
  couplings to matter fields in scalar-tensor gravity},'' {\em Phys. Rev. D}
  {\bf 108} (2023), no.~4 044006, \href{https://arxiv.org/abs/2302.03418}{{\tt
  2302.03418}}.

\bibitem{Arkani-Hamed:2003pdi}
N.~Arkani-Hamed, H.-C. Cheng, M.~A. Luty, and S.~Mukohyama, ``{Ghost
  condensation and a consistent infrared modification of gravity},'' {\em JHEP}
  {\bf 05} (2004) 074, \href{https://arxiv.org/abs/hep-th/0312099}{{\tt
  hep-th/0312099}}.

\bibitem{Arkani-Hamed:2003juy}
N.~Arkani-Hamed, P.~Creminelli, S.~Mukohyama, and M.~Zaldarriaga, ``{Ghost
  inflation},'' {\em JCAP} {\bf 04} (2004) 001,
  \href{https://arxiv.org/abs/hep-th/0312100}{{\tt hep-th/0312100}}.

\bibitem{Cheung:2007st}
C.~Cheung, P.~Creminelli, A.~L. Fitzpatrick, J.~Kaplan, and L.~Senatore, ``{The
  Effective Field Theory of Inflation},'' {\em JHEP} {\bf 03} (2008) 014,
  \href{https://arxiv.org/abs/0709.0293}{{\tt 0709.0293}}.

\bibitem{Gubitosi:2012hu}
G.~Gubitosi, F.~Piazza, and F.~Vernizzi, ``{The Effective Field Theory of Dark
  Energy},'' {\em JCAP} {\bf 02} (2013) 032,
  \href{https://arxiv.org/abs/1210.0201}{{\tt 1210.0201}}.

\bibitem{Noller:2019chl}
J.~Noller, L.~Santoni, E.~Trincherini, and L.~G. Trombetta, ``{Black Hole
  Ringdown as a Probe for Dark Energy},'' {\em Phys. Rev. D} {\bf 101} (2020)
  084049, \href{https://arxiv.org/abs/1911.11671}{{\tt 1911.11671}}.

\bibitem{Mukohyama:2022enj}
S.~Mukohyama and V.~Yingcharoenrat, ``{Effective field theory of black hole
  perturbations with timelike scalar profile: formulation},'' {\em JCAP} {\bf
  09} (2022) 010, \href{https://arxiv.org/abs/2204.00228}{{\tt 2204.00228}}.

\bibitem{Franciolini:2018uyq}
G.~Franciolini, L.~Hui, R.~Penco, L.~Santoni, and E.~Trincherini, ``{Effective
  Field Theory of Black Hole Quasinormal Modes in Scalar-Tensor Theories},''
  {\em JHEP} {\bf 02} (2019) 127, \href{https://arxiv.org/abs/1810.07706}{{\tt
  1810.07706}}.

\bibitem{Hui:2021cpm}
L.~Hui, A.~Podo, L.~Santoni, and E.~Trincherini, ``{Effective Field Theory for
  the perturbations of a slowly rotating black hole},'' {\em JHEP} {\bf 12}
  (2021) 183, \href{https://arxiv.org/abs/2111.02072}{{\tt 2111.02072}}.

\bibitem{Khoury:2022zor}
J.~Khoury, T.~Noumi, M.~Trodden, and S.~S.~C. Wong, ``{Stability of hairy black
  holes in shift-symmetric scalar-tensor theories via the effective field
  theory approach},'' {\em JCAP} {\bf 04} (2023) 035,
  \href{https://arxiv.org/abs/2208.02823}{{\tt 2208.02823}}.

\bibitem{LIGOScientific:2016aoc}
{\bf LIGO Scientific, Virgo} Collaboration, B.~P. Abbott {\em et~al.},
  ``{Observation of Gravitational Waves from a Binary Black Hole Merger},''
  {\em Phys. Rev. Lett.} {\bf 116} (2016), no.~6 061102,
  \href{https://arxiv.org/abs/1602.03837}{{\tt 1602.03837}}.

\bibitem{LIGOScientific:2018mvr}
{\bf LIGO Scientific, Virgo} Collaboration, B.~P. Abbott {\em et~al.},
  ``{GWTC-1: A Gravitational-Wave Transient Catalog of Compact Binary Mergers
  Observed by LIGO and Virgo during the First and Second Observing Runs},''
  {\em Phys. Rev. X} {\bf 9} (2019), no.~3 031040,
  \href{https://arxiv.org/abs/1811.12907}{{\tt 1811.12907}}.

\bibitem{LIGOScientific:2019lzm}
{\bf LIGO Scientific, Virgo} Collaboration, R.~Abbott {\em et~al.}, ``{Open
  data from the first and second observing runs of Advanced LIGO and Advanced
  Virgo},'' {\em SoftwareX} {\bf 13} (2021) 100658,
  \href{https://arxiv.org/abs/1912.11716}{{\tt 1912.11716}}.

\bibitem{KAGRA:2023pio}
{\bf KAGRA, VIRGO, LIGO Scientific} Collaboration, R.~Abbott {\em et~al.},
  ``{Open Data from the Third Observing Run of LIGO, Virgo, KAGRA, and GEO},''
  {\em Astrophys. J. Suppl.} {\bf 267} (2023), no.~2 29,
  \href{https://arxiv.org/abs/2302.03676}{{\tt 2302.03676}}.

\bibitem{Mukohyama:2022skk}
S.~Mukohyama, K.~Takahashi, and V.~Yingcharoenrat, ``{Generalized Regge-Wheeler
  equation from Effective Field Theory of black hole perturbations with a
  timelike scalar profile},'' {\em JCAP} {\bf 10} (2022) 050,
  \href{https://arxiv.org/abs/2208.02943}{{\tt 2208.02943}}.

\bibitem{Mukohyama:2023xyf}
S.~Mukohyama, K.~Takahashi, K.~Tomikawa, and V.~Yingcharoenrat, ``{Quasinormal
  modes from EFT of black hole perturbations with timelike scalar profile},''
  {\em JCAP} {\bf 07} (2023) 050, \href{https://arxiv.org/abs/2304.14304}{{\tt
  2304.14304}}.

\bibitem{Konoplya:2023ppx}
R.~A. Konoplya, ``{Quasinormal modes and grey-body factors of regular black
  holes with a scalar hair from the Effective Field Theory},'' {\em JCAP} {\bf
  07} (2023) 001, \href{https://arxiv.org/abs/2305.09187}{{\tt 2305.09187}}.

\bibitem{Oshita:2024fzf}
N.~Oshita, K.~Takahashi, and S.~Mukohyama, ``{(In)stability of the black hole
  greybody factors and ringdowns against a small-bump correction},''
  \href{https://arxiv.org/abs/2406.04525}{{\tt 2406.04525}}.

\bibitem{Barura:2024uog}
C.~G.~A. Barura, H.~Kobayashi, S.~Mukohyama, N.~Oshita, K.~Takahashi, and
  V.~Yingcharoenrat, ``{Tidal Love Numbers from EFT of Black Hole Perturbations
  with Timelike Scalar Profile},'' \href{https://arxiv.org/abs/2405.10813}{{\tt
  2405.10813}}.

\bibitem{Aoki:2021wew}
K.~Aoki, M.~A. Gorji, S.~Mukohyama, and K.~Takahashi, ``{The effective field
  theory of vector-tensor theories},'' {\em JCAP} {\bf 01} (2022), no.~01 059,
  \href{https://arxiv.org/abs/2111.08119}{{\tt 2111.08119}}.

\bibitem{Aoki:2023bmz}
K.~Aoki, M.~A. Gorji, S.~Mukohyama, K.~Takahashi, and V.~Yingcharoenrat,
  ``{Effective field theory of black hole perturbations in vector-tensor
  gravity},'' {\em JCAP} {\bf 03} (2024) 012,
  \href{https://arxiv.org/abs/2311.06767}{{\tt 2311.06767}}.

\bibitem{Piazza:2013pua}
F.~Piazza, H.~Steigerwald, and C.~Marinoni, ``{Phenomenology of dark energy:
  exploring the space of theories with future redshift surveys},'' {\em JCAP}
  {\bf 1405} (2014) 043, \href{https://arxiv.org/abs/1312.6111}{{\tt
  1312.6111}}.

\bibitem{Traykova:2019oyx}
D.~Traykova, E.~Bellini, and P.~G. Ferreira, ``{The phenomenology of beyond
  Horndeski gravity},'' {\em JCAP} {\bf 08} (2019) 035,
  \href{https://arxiv.org/abs/1902.10687}{{\tt 1902.10687}}.

\bibitem{Peirone:2019yjs}
S.~Peirone, G.~Benevento, N.~Frusciante, and S.~Tsujikawa, ``{Cosmological
  constraints and phenomenology of a beyond-Horndeski model},'' {\em Phys. Rev.
  D} {\bf 100} (2019), no.~6 063509,
  \href{https://arxiv.org/abs/1905.11364}{{\tt 1905.11364}}.

\bibitem{Creminelli:2017sry}
P.~Creminelli and F.~Vernizzi, ``{Dark Energy after GW170817 and GRB170817A},''
  {\em Phys. Rev. Lett.} {\bf 119} (2017), no.~25 251302,
  \href{https://arxiv.org/abs/1710.05877}{{\tt 1710.05877}}.

\bibitem{Creminelli:2018xsv}
P.~Creminelli, M.~Lewandowski, G.~Tambalo, and F.~Vernizzi, ``{Gravitational
  Wave Decay into Dark Energy},'' {\em JCAP} {\bf 1812} (2018), no.~12 025,
  \href{https://arxiv.org/abs/1809.03484}{{\tt 1809.03484}}.

\bibitem{Creminelli:2019nok}
P.~Creminelli, G.~Tambalo, F.~Vernizzi, and V.~Yingcharoenrat, ``{Resonant
  Decay of Gravitational Waves into Dark Energy},'' {\em JCAP} {\bf 10} (2019)
  072, \href{https://arxiv.org/abs/1906.07015}{{\tt 1906.07015}}.

\bibitem{Creminelli:2019kjy}
P.~Creminelli, G.~Tambalo, F.~Vernizzi, and V.~Yingcharoenrat, ``{Dark-Energy
  Instabilities induced by Gravitational Waves},'' {\em JCAP} {\bf 05} (2020)
  002, \href{https://arxiv.org/abs/1910.14035}{{\tt 1910.14035}}.

\bibitem{Domenech:2015tca}
G.~Dom\`enech, S.~Mukohyama, R.~Namba, A.~Naruko, R.~Saitou, and Y.~Watanabe,
  ``{Derivative-dependent metric transformation and physical degrees of
  freedom},'' {\em Phys. Rev. D} {\bf 92} (2015), no.~8 084027,
  \href{https://arxiv.org/abs/1507.05390}{{\tt 1507.05390}}.

\bibitem{Takahashi:2017zgr}
K.~Takahashi, H.~Motohashi, T.~Suyama, and T.~Kobayashi, ``{General invertible
  transformation and physical degrees of freedom},'' {\em Phys. Rev. D} {\bf
  95} (2017), no.~8 084053, \href{https://arxiv.org/abs/1702.01849}{{\tt
  1702.01849}}.

\bibitem{Motohashi:2016prk}
H.~Motohashi, T.~Suyama, and K.~Takahashi, ``{Fundamental theorem on gauge
  fixing at the action level},'' {\em Phys. Rev. D} {\bf 94} (2016), no.~12
  124021, \href{https://arxiv.org/abs/1608.00071}{{\tt 1608.00071}}.

\bibitem{Hayward:2005gi}
S.~A. Hayward, ``{Formation and evaporation of regular black holes},'' {\em
  Phys. Rev. Lett.} {\bf 96} (2006) 031103,
  \href{https://arxiv.org/abs/gr-qc/0506126}{{\tt gr-qc/0506126}}.

\bibitem{Takahashi:2019oxz}
K.~Takahashi, H.~Motohashi, and M.~Minamitsuji, ``{Linear stability analysis of
  hairy black holes in quadratic degenerate higher-order scalar-tensor
  theories: Odd-parity perturbations},'' {\em Phys. Rev. D} {\bf 100} (2019),
  no.~2 024041, \href{https://arxiv.org/abs/1904.03554}{{\tt 1904.03554}}.

\bibitem{BenAchour:2020wiw}
J.~Ben~Achour, H.~Liu, and S.~Mukohyama, ``{Hairy black holes in DHOST
  theories: Exploring disformal transformation as a solution-generating
  method},'' {\em JCAP} {\bf 02} (2020) 023,
  \href{https://arxiv.org/abs/1910.11017}{{\tt 1910.11017}}.

\bibitem{Gleyzes:2013ooa}
J.~Gleyzes, D.~Langlois, F.~Piazza, and F.~Vernizzi, ``{Essential Building
  Blocks of Dark Energy},'' {\em JCAP} {\bf 08} (2013) 025,
  \href{https://arxiv.org/abs/1304.4840}{{\tt 1304.4840}}.

\end{thebibliography}\endgroup

\end{document}